\input harvmac
\input epsf.sty

\def\tx{{\tilde x}}
\def\ty{{\tilde y}}
\def\tz{{\tilde z}}
\def\tw{{\tilde w}}
\lref\AV{
  M.~Aganagic and C.~Vafa,
  ``G(2) manifolds, mirror symmetry and geometric engineering,''
  arXiv:hep-th/0110171.
}
\lref\ABJM{
  O.~Aharony, O.~Bergman, D.~L.~Jafferis and J.~Maldacena,
  ``N=6 superconformal Chern-Simons-matter theories, M2-branes and their
  gravity duals,''
  JHEP {\bf 0810}, 091 (2008)
  [arXiv:0806.1218 [hep-th]].
}
\lref\Denef{
  F.~Denef,
  ``Les Houches Lectures on Constructing String Vacua,''
  arXiv:0803.1194 [hep-th].
}
\lref\HZ{
  A.~Hanany and A.~Zaffaroni,
  ``Tilings, Chern-Simons Theories and M2 Branes,''
  JHEP {\bf 0810}, 111 (2008)
  [arXiv:0808.1244 [hep-th]].
}
\lref\HVZ{
  A.~Hanany, D.~Vegh and A.~Zaffaroni,
  ``Brane Tilings and M2 Branes,''
  JHEP {\bf 0903}, 012 (2009)
  [arXiv:0809.1440 [hep-th]].
}
\lref\hv{
  K.~Hori, A.~Iqbal and C.~Vafa,
  ``D-branes and mirror symmetry,''
  arXiv:hep-th/0005247.
}
\lref\witten{
  E.~Witten,
  ``Phases of N = 2 theories in two dimensions,''
  Nucl.\ Phys.\  B {\bf 403}, 159 (1993)
  [arXiv:hep-th/9301042].
}
\lref\Shamit{
  S.~Kachru and D.~Simic,
  ``Stringy Instantons in IIB Brane Systems,''
  arXiv:0803.2514 [hep-th].
}
\lref\HananyYHH{
  A.~Hanany and Y.~H.~He,
  ``M2-Branes and Quiver Chern-Simons: A Taxonomic Study,''
  arXiv:0811.4044 [hep-th].
}
\lref\FHPR{
  S.~Franco, A.~Hanany, J.~Park and D.~Rodriguez-Gomez,
  ``Towards M2-brane Theories for Generic Toric Singularities,''
  JHEP {\bf 0812}, 110 (2008)
  [arXiv:0809.3237 [hep-th]].
}
\lref\DM{
  M.~R.~Douglas and G.~W.~Moore,
  ``D-branes, Quivers, and ALE Instantons,''
  arXiv:hep-th/9603167.
}
\lref\orbifolds{
  M.~R.~Douglas, B.~R.~Greene and D.~R.~Morrison,
  ``Orbifold resolution by D-branes,''
  Nucl.\ Phys.\  B {\bf 506}, 84 (1997)
  [arXiv:hep-th/9704151].
}
\lref\MSI{
  D.~Martelli and J.~Sparks,
  ``Moduli spaces of Chern-Simons quiver gauge theories and AdS(4)/CFT(3),''
  Phys.\ Rev.\  D {\bf 78}, 126005 (2008)
  [arXiv:0808.0912 [hep-th]].
}
\lref\MSII{
  D.~Martelli and J.~Sparks,
  ``Notes on toric Sasaki-Einstein seven-manifolds and $AdS_4/CFT_3$,''
  JHEP {\bf 0811}, 016 (2008)
  [arXiv:0808.0904 [hep-th]].
}
\lref\JA{
  D.~L.~Jafferis and A.~Tomasiello,
  ``A simple class of N=3 gauge/gravity duals,''
  JHEP {\bf 0810}, 101 (2008)
  [arXiv:0808.0864 [hep-th]].
}
\lref\UY{
  K.~Ueda and M.~Yamazaki,
  ``Toric Calabi-Yau four-folds dual to Chern-Simons-matter theories,''
  JHEP {\bf 0812}, 045 (2008)
  [arXiv:0808.3768 [hep-th]].
}
\lref\GTI{
  D.~Gaiotto and A.~Tomasiello,
  ``The gauge dual of Romans mass,''
  arXiv:0901.0969 [hep-th].
}
\lref\GTII{
  D.~Gaiotto and A.~Tomasiello,
  ``Perturbing gauge/gravity duals by a Romans mass,''
  arXiv:0904.3959 [hep-th].
}
\lref\hp{
  J.~Davey, A.~Hanany, N.~Mekareeya and G.~Torri,
  ``Phases of M2-brane Theories,''
  arXiv:0903.3234 [hep-th].
}
\lref\feng{
  B.~Feng, A.~Hanany and Y.~H.~He,
  ``D-brane gauge theories from toric singularities and toric duality,''
  Nucl.\ Phys.\  B {\bf 595}, 165 (2001)
  [arXiv:hep-th/0003085].
  B.~Feng, A.~Hanany and Y.~H.~He,
  ``Phase structure of D-brane gauge theories and toric duality,''
  JHEP {\bf 0108}, 040 (2001)
  [arXiv:hep-th/0104259].
}
{\Title{\vbox{
\hbox{%
}
}}
{\vbox{
\centerline{A Stringy Origin}
\vskip .1in
\centerline{of}
\vskip .1in
\centerline{M2 Brane Chern-Simons Theories}}}
\vskip .3in
\centerline{Mina Aganagic}
\vskip .2in
}

\centerline{Center for Theoretical Physics, University of California, Berkeley, CA 94720}

\vskip .4in

We show that string duality relates M-theory on a local Calabi-Yau fourfold singularity $X_4$
to type IIA string theory on a Calabi-Yau threefold $X_3$ fibered over a real line, with
RR 2-form fluxes turned on. The RR flux encodes how the M-theory circle is fibered over the
IIA geometry. The theories on $N$ D2 branes probing $X_3$ are the well-known quiver theories with ${\cal N}=2$ supersymmetry in three dimensions. We show that turning on fluxes, and fibering the $X_3$
over a direction transverse to the branes, corresponds to turning on ${\cal N}=2$ Chern-Simons couplings. String duality implies that, in the strong coupling limit, the $N$ D2 branes on $X_3$ in this background become $N$ M2 branes on $X_4$. This provides a string theory derivation for the recently conjectured description of the M2 brane theories on Calabi-Yau fourfolds in terms of ${\cal N}=2$ quiver Chern-Simons theories. We also provide a new ${\cal N}=2$ Chern-Simons theory dual to $AdS_4 \times Q^{1,1,1}$. Type IIA/M-theory duality also relates
IIA string theory on $X_3$ with only the RR fluxes turned on,
to M-theory on a $G_2$ holonomy manifold. We show that this implies that the $N$ M2 branes probing the $G_2$ manifold are described by the quiver Chern-Simons theory originating from the D2 branes probing $X_3$, except that now Chern-Simons terms preserve only ${\cal N}=1$ supersymmetry in three dimensions.
\vfill
\eject

\newsec{Introduction}

Duality provides powerful tools to understand string theory.
AdS/CFT correspondence defines quantum gravity in
AdS backgrounds in terms of the dual conformal field theory. At the same time,
gravity on AdS space describes the conformal theory at large $N$.
Similarly IIA/M-theory duality provides a way to understand the strong coupling limit of type IIA
string theory in terms of 11 dimensional supergravity. Conversely, in compactifications of M-theory on a circle, IIA string theory defines the still mysterious quantum theory that underlies M-theory.

One of the mysteries of M-theory is how to describe the theory on $N$ coincident M2 branes.
At low energies, this should be the conformal field theory dual to M-theory
on $AdS_4\times M_7$ where $M_7$ is seven dimensional Einstein manifold \ref\DuffHR{
  M.~J.~Duff, B.~E.~W.~Nilsson and C.~N.~Pope,
  ``Kaluza-Klein Supergravity,''
  Phys.\ Rept.\  {\bf 130}, 1 (1986).
}.
The theories on $N$ M2 branes in M-theory are dual to the theory on $N$ D2 branes in IIA.
The latter theory is a gauge theory, with the dimensionfull coupling constant proportional to $g_s$.
In the strong coupling limit, the D2 brane theory should describe M2 branes.
However, the resulting theory is apparently always strongly coupled, so the description does not seem useful

A way around this was proposed in \ref\Schwarz{J.~H.~Schwarz,
  ``Superconformal Chern-Simons theories,''
  JHEP {\bf 0411}, 078 (2004)
  [arXiv:hep-th/0411077].
}. The idea is that, at low energies, the theory on the M2 branes may be a Chern-Simons theory, with matter. The Chern-Simons level provides a dimensionless coupling constant needed to define the gauge theory.
Moreover, Chern-Simons interaction is scale invariant, being independent of the metric, so the
theory could also be conformal. Recently, \ABJM\ constructed a Chern-Simons theory corresponding to $N$ M2 branes on Calabi-Yau fourfold $X_4 ={\bf C}^4/{\bf Z}_k$, where $k$ gets identified with the level. This theory is conformal, and dual, at large $N$ to M-theory on $AdS_4 \times S^7/{\bf Z}_k$. Moreover, for $k=1$ it should correspond to M2 branes on flat space, with maximal supersymmetry in three dimensions.  Various authors \refs{\MSI,\HZ,\JA, \HVZ, \UY} proposed generalizations of this construction to Chern-Simons theories describing M2 branes on local Calabi-Yau fourfolds. A check of these proposals is that for a single M2 brane, the moduli space is indeed the Calabi-Yau fourfold $X_4$. However, in general, there is more than one such Chern-Simons theory that one can write down which would still give the same moduli space \refs{\FHPR,\hp, \HananyYHH}. It is possible that there are dualities relating some of these theories, but we cannot a-priori exclude that they flow to different theories at long distance.

The aim of this paper is to provide a stringy derivation of the Chern-Simons theory on
$N$ M2 branes probing a local Calabi-Yau fourfold $X_4$\foot{Such a derivation was proposed in \refs{\ABJM,\JA} for the theories studied there. While the string realizations are dual to each other, our proposal is simpler, and the relevant string backgrounds are better understood. Moreover, we provide a unified derivation, applicable for a large class of toric Calabi-Yau singularities.}. To begin with, we show that, M-theory on a local toric Calabi-Yau fourfold $X_4$ is dual to IIA string theory on
a Calabi-Yau threefold $X_3$ fibered over ${\bf R}$. The Kahler moduli of $X_3$
vary over ${\bf R}$ and moreover, there are RR 2-form fluxes on $X_3\times {\bf R}$.
This is a consequence of a purely geometrical observation: The toric Calabi-Yau fourfold $X_4$ is a
is a circle fibration over $X_3 \times {\bf R}$. Moreover, the viewing the coordinate on the $S^1$ as a
section of a $U(1)$ bundle on $X_3 \times {\bf R}$ the curvature of the corresponding connection is easy to compute. When we compactify M-theory on $X_4$, the IIA/M theory duality relates it to IIA on $X_3 \times {\bf R}$, where the curvature of the $U(1)$ bundle gets identified with the RR 2-form flux in IIA.

The theory on $N$ D2 branes in IIA probing the local Calabi-Yau threefold $X_3$ is well known, in the limit where $X_3$ develops a singularity. These are ${\cal N}=2$ quiver gauge theories in three dimensions, with superpotential. More precisely, the theories on Dp branes probing $X_3$ were studied extensively for D3 branes and D0 branes\foot{The former were studied as examples of conformal field theories in four dimensions with four supercharges (see for example \ref\GUD{
  F.~Cachazo, B.~Fiol, K.~A.~Intriligator, S.~Katz and C.~Vafa,
  ``A geometric unification of dualities,''
  Nucl.\ Phys.\  B {\bf 628}, 3 (2002)
  [arXiv:hep-th/0110028].
}\ref\FrancoRJ{
  S.~Franco, A.~Hanany, K.~D.~Kennaway, D.~Vegh and B.~Wecht,
  ``Brane Dimers and Quiver Gauge Theories,''
  JHEP {\bf 0601}, 096 (2006)
  [arXiv:hep-th/0504110].
}\ref\FrancoSM{
  S.~Franco, A.~Hanany, D.~Martelli, J.~Sparks, D.~Vegh and B.~Wecht,
  ``Gauge theories from toric geometry and brane tilings,''
  JHEP {\bf 0601}, 128 (2006)
  [arXiv:hep-th/0505211].
}\ref\KV{
  B.~Feng, Y.~H.~He, K.~D.~Kennaway and C.~Vafa,
  ``Dimer models from mirror symmetry and quivering amoebae,''
  Adv.\ Theor.\ Math.\ Phys.\  {\bf 12}, 3 (2008)
  [arXiv:hep-th/0511287].
}) while the latter were studied in the context of 4d ${\cal N}=2$ black holes (see for example, \ref\DenefRU{
  F.~Denef,
  ``Quantum quivers and Hall/hole halos,''
  JHEP {\bf 0210}, 023 (2002)
  [arXiv:hep-th/0206072].
}\ref\OoguriYB{
  H.~Ooguri and M.~Yamazaki,
  ``Crystal Melting and Toric Calabi-Yau Manifolds,''
  arXiv:0811.2801 [hep-th].
}). Detailed derivation of the quiver theory from mirror symmetry \hv\ and the classical geometry of intersecting three-cycles was presented in \GUD .
An alternative derivation was developed in \feng\ and many followup works, starting from the theories of branes on orbifolds
${\bf C}^3/{\bf Z}_k\times {\bf Z_m}$, which have a free worldsheet CFT description and following their deformations.}. However, given $X_3$, the theories on $N$ D-branes probing it are the same, at least classically, in any dimension -- differing only by compactifications on tori, and dimensional reduction.
This is because the theory on the branes is determined by internal CFT on a disk -- and the answers can be formulated in terms of the geometry of $X_3$ and fractional branes wrapping it.

Next, we consider deforming the theory by turning on RR fluxes and fibering $X_3$ over ${\bf R}$,
so that the type IIA string theory becomes dual to M-theory on a local Calabi-Yau fourfold singularity $X_4$.
On the one hand, the deformation has to preserve ${\cal N}=2$ supersymmetry on the D2 brane world volume, since the theory lifts to M2 branes on the Calabi-Yau fourfold. On the other hand, turning on RR fluxes on $X_3$ corresponds to turning on Chern-Simons couplings in the gauge theory. Namely, at the Calabi-Yau singularity, the D2 branes split into fractional branes, wrapped on the vanishing cycles. The Wess-Zumino terms on the brane world-volume, in the presence of RR flux, induce Chern-Simons couplings in three dimensions. By supersymmetry then, these become the ${\cal N}=2$ preserving Chern-Simons terms.
This is exactly what is needed to describe D2 branes on $X_3$ fibered over ${\bf R}$.
Namely, ${\cal N}=2$ Chern-Simons terms imply that Fayet-Iliopoulos terms vary as a function of the real scalar field in the ${\cal N}=2$ vector multiplet. The later parameterizes the direction in ${\bf R}^{3,1}$ transverse to the D2 branes. Since Fayet-Iliopoulos terms in the gauge theory correspond to Kahler moduli in the Calabi-Yau, this is exactly what is needed
to describe $N$ D2 branes on $X_3$ fibered over ${\bf R}$.

Consequently, string duality implies that ${\cal N}=2$ quiver theory on $N$ D2 branes probing $X_3$, with ${\cal N}=2$ Chern-Simons terms turned on lifts to the theory on $N$ M2 branes probing $X_4$.
At strong coupling, the gauge kinetic terms vanish, and the theory becomes a Chern-Simons quiver theory.
These are precisely the theories proposed in \refs{\ABJM,\JA,\MSI,\HZ, \HVZ, \UY} to describe $N$ M2 branes on various local Calabi-Yau fourfold singularities. The relation to the theories on the D3 branes probing $X_3$ was noticed there, but its physical meaning was missed.
As mentioned above, one test of the proposals in the literature is that the moduli space of a single M2 brane should be $X_4$. What we are proposing here is stronger, since the theories on D2 branes
probing $X_3\times {\bf R}$, with fluxes, can be {\it derived from string theory}, as we explained.
This should be contrasted with the theories "without 4d parents" in \refs{\HananyYHH, \FHPR, \hp}\foot{Some of these theories do pass some stringent tests. See for example \ref\klebanov{
  S.~Franco, I.~R.~Klebanov and D.~Rodriguez-Gomez,
  ``M2-branes on Orbifolds of the Cone over $Q^{1,1,1},$''
  arXiv:0903.3231 [hep-th].
}, \hp.}.

It is natural to expect that the quiver Chern-Simons theory of $X_4$ we wrote down is a conformal field theory in three dimensions. This has already been shown in cases studied
in \refs{\ABJM,\JA}. It should hold more generally, since on general grounds,
local Calabi-Yau fourfold singularities $X_4$'s should be cones over Einstein-Sasaki 7-manifolds $Y_7$ \ref\acharya{
  B.~S.~Acharya, J.~M.~Figueroa-O'Farrill, C.~M.~Hull and B.~J.~Spence,
  ``Branes at conical singularities and holography,''
  Adv.\ Theor.\ Math.\ Phys.\  {\bf 2}, 1249 (1999)
  [arXiv:hep-th/9808014].
}. We have shown that the quiver Chern-Simons theory describes $N$ M2 branes at the tip of this cone. In the near horizon limit, the branes deform the geometry to $AdS_4 \times Y_7,$ so at least in the large $N$ limit, the conformal symmetry of the theory is manifest\foot{The cases where the explicit cone metrics are known are limited. See for example \MSII\ and references therein.}.

Finally, similar ideas can be used to find the theories on $N$ M2 branes probing local $G_2$ holonomy manifolds\foot{The Chern-Simons theories on $1$ M2 brane probing a local $G_2$ holonomy manifold were first discovered in \AV .}. Namely IIA/M-theory duality relates IIA on a Calabi-Yau threefold $X_3$ with RR fluxes turned on, to M-theory on a $G_2$ holonomy manifold. Turning on fluxes in this case corresponds to deforming the ${\cal N}=2$ quiver gauge theories on $N$ D2 branes probing $X_3$ by ${\cal N}=1$ Chern-Simons terms. At low energies, these theories become the theories on $N$ M2 branes on the corresponding $G_2$ holonomy manifolds. Since the gauge kinetic terms vanish in this limit,
the theories describing the M2 branes are ${\cal N}=1$ quiver Chern-Simons theories. We do not expect these theories to be conformal, as $G_2 \times {\bf R}$ is not a cone, and no corresponding Einstein 7-manifolds are known \acharya .

The paper is organized as follows.
In section 2 we show that a toric Calabi-Yau fourfold can be thought of as a circle fibration over $X_3 \times {\bf R}.$ We show how to find the first Chern class of the fibration. This implies that  M-theory compactified on $X_4$ is dual to IIA on $X_3 \times {\bf R}$ with fluxes turned on.
In section 3 we derive from string theory the theories on $N$ D2 branes probing $X_3$, in these backgrounds. For one brane, we show that the moduli space becomes $X_4$, similarly to \refs{\HZ,\HVZ, \DM}. Moreover, $X_4$ is fibered over $X_3 \times {\bf R}$ in a manner consistent with the Chern-Simons terms turned on. Namely, in IIA/M-theory duality the center of mass $U(1)$ gauge field on the D2 brane is dual, via scalar vector duality in three dimensions, to the compact scalar field parameterizing the M2 brane position on the M-theory circle. We will show that in the presence of Chern-Simons terms, this scalar picks up a charge. This has the interpretation of having the M-theory circle be fibered non-trivially over $X_3$.
Moreover, we will show that the Chern class of this fibration, is precisely such to correspond to RR 2-form fluxes that generated the Chern-Simons terms on the D2 brane in the first place.
In sections 4-8 we give some examples. The examples of sections 4-7 have already appeared in literature. We are providing a string theory derivation of the corresponding Chern-Simons quiver theories and an interpretation, in the context of IIA/M-theory duality. The example in section 8 is new, providing a new quiver Chern-Simons theory that should be dual, based on IIA/M-theory duality to $N$ M2 branes on a cone over $Q^{1,1,1}$.\foot{This theory was discussed recently in \ref\AmaritiRB{
  A.~Amariti, D.~Forcella, L.~Girardello and A.~Mariotti,
  ``3D Seiberg-like Dualities and M2 Branes,''
  arXiv:0903.3222 [hep-th].
}, but this connection was not made.}
Finally, in section 9 we use IIA/M-theory duality to derive the ${\cal N}=1$ quiver
Chern-Simons theories on M2 branes probing local $G_2$ holonomy manifolds.

\newsec{Calabi-Yau Fourfolds and M-theory/IIA duality}

In this section from purely geometric considerations, we will derive a duality between M-theory compactification on a Calabi-Yau fourfold $X_4$ and type IIA on a Calabi-Yau threefold $X_3$ fibered over a real line ${\bf R}$, with fluxes turned on. Namely, we will show here that a Calabi-Yau fourfold $X_4$ can be described as a circle fibration over a base. The base is a Calabi-Yau 3-fold $X_3$ fibred  over a real line ${\bf R}$. The first Chern Class of the circle fibration is typically non-vanishing. This implies that when we compactify M-theory on $X_4$, there will be RR 2-form fluxes turned on, on $X_3 \times {\bf R}$. Later on, we will argue that the string duality then implies that M2 brane probes of $X_4$ are dual to D2 brane probes of the IIA geometry, in the strong coupling limit of the later.

\subsec{Geometric considerations}

A convenient way to describe the Calabi-Yau geometry \witten\ is as a moduli space of a linear sigma model. To describe a $d$ dimensional Calabi-Yau manifold $X_d$ we start with a copy of ${\bf C}^{N+d}$,
for some $N$, parameterized by complex coordinates
$$
\phi_i, \qquad i=1,\ldots N+d,
$$
Next, for each $\phi_i$, we pick a set of charges $Q^i_a$, for $a=1, \ldots , N$
satisfying
\eqn\CY{
\sum_{i} Q_i^a = 0.
}
The Calabi-Yau manifold is than obtained by setting
\eqn\Dflat{
\sum_i Q_i^a \, |\phi_i|^2 = r^a,
}
and dividing by the gauge group
\eqn\gauge{
\phi_i \rightarrow \phi_i \,\exp(i\, \lambda_a Q_a^i).
}
The parameters $r^a$ correspond to Kahler moduli of the Calabi-Yau manifold.
Mathematically, the above construction is called a symplectic quotient.
As explained in \witten , symplectic quotient construction
construction can be realized physically as a Higgs branch of a $(2,2)$ supersymmetric linear sigma model in two dimensions, with
$\phi_i$ corresponding to the bottom components of chiral fields $\Phi_i$, and with gauge group $U(1)^N$. Closer to our purposes, it can also be interpreted in terms of a theory in one dimension higher, i.e.
in terms of a Higgs branch of an abelian gauge theory in three dimensions, with four supercharges.

We will now show that we can view the $d$ dimensional Calabi-Yau threefold $X_d$ as an $S^1$ fibration over a base, which itself is a fibration of a $d-1$ dimensional Calabi-Yau $X_{d-1}$ over a line ${\bf R}$.
To see this, consider adding
a complex variable
$$
r_0+i\, \theta_0
$$
and pick a set of charges $Q_i^0$, which also satisfy the Calabi-Yau condition,
$$\sum_i Q_i^0 =0.
$$
To avoid changing the manifold, we introduce an additional constraint
\eqn\Dflato{
\sum_i Q_i^0\, |\phi_i| = r_0
}
(note that $r_0$ is variable now) and an additional gauge symmetry
\eqn\gaugeo{\eqalign{
\theta_0 &\rightarrow \theta_0 + \lambda\cr
\phi_i &\rightarrow  \exp(i Q_i^0 \lambda) \, \phi_i.
}}
Namely, using \Dflato\ we can solve for $r_0$ and gauge away $\theta_0$ using
\gaugeo . So, as expected, we have just written the original Calabi-Yau $X_d$ in a different way\foot{We could also consider replacing $r_0$ by $k r_0$ while giving $\theta_0$ charge $k$ in \gaugeo , corresponding to sending $\theta_0$ to $\theta_0 + k \lambda$ instead. In this case we do change the manifold, from $X_d$ to a ${\bf Z}_k$ quotient $X_d/{\bf Z}_k$, since gauge transformations by $\lambda = 2\pi i/k$ leave $\theta_0$ invariant.}.

However, this gives us a way to
think about $X_d$ as a circle fibration. Namely, the $S^1$ fiber of the Calabi-Yau is $\theta_0$. To find the base, we consider a projection that "forgets" $\theta_0$.
At {\it fixed} $r_0$, the base is simply a $d-1$ dimensional Calabi-Yau manifold $X_{d-1}$, given by solving \Dflat\Dflato\
\eqn\df{\eqalign{
\sum_i Q_i^a \,&|\phi_i|^2 = r^a,\cr
\sum_i Q_i^0 \, &|\phi_i| = r^0}
}
and dividing by $N+1$ $U(1)$ gauge transformations \gauge\ and  \gaugeo .
Allowing $r_0$ to vary, the lower dimensional Calabi-Yau manifold $X_{d-1}$ is a fibration over the
real line parameterized by $r_0$, where as we vary $r_0$, the Kahler moduli of $X_{d-1}$ vary according to \df .

We can easily find the first Chern class of the fibration. To begin with, note that
$$
d\theta_0
$$
is not gauge invariant because $d \theta_0$ transforms under \gaugeo . The gauge invariant combination is
$$
d \theta_0 + \sum_j iq_j  \, d\phi_j/ \phi_j
$$
where
$$
\sum_j q_j Q_j^0 = 1, \qquad \qquad \sum_j q_j Q_j^a = 0, \qquad a=1, \ldots ,N
$$
Then $A_{RR} =  \sum_j \,q_j  \, d\phi_j/ \phi_j$ is the connection on the circle bundle over $X_3 \times {\bf R}$ whose curvature
is
\eqn\fluxo{
F_{RR} = dA_{RR} = \sum_j q_j \omega_{j}
}
where $\omega_j = \delta^2(\phi_j) d \phi_j \wedge d {\overline \phi_j}$ is a two-form with $\delta$-function support on the divisor $D_j$ corresponding to setting $\phi_j=0$. We can write
\eqn\fluxa{
[F_{RR}] = \sum_j q_j\; [D_j],
}
for the cohomology classes $[F_{RR}]$ and $[D_i]$ of $F_{RR}$ and $\omega_i$, respectively.
Consider now the flux of $F_{RR}$ through a compact 2-cycle $C$ in the geometry,
$$\int_{C} F_{RR}.
$$
There is a correspondence between generators of $H_2(X_{d-1})$ and the $U(1)$ gauge groups in the linear sigma model. For a curve $C$, let
$$
\sum_i Q_i^C |\phi_i|^2 = r_C
$$
denote the corresponding D-term. In particular, $r_C$ is the Kahler class of the curve. It is a standard result in toric geometry (see for example \Denef\ for a recent review) that
$$
\# (D_j \cap C) = Q_j^C,
$$
and using this, we can evaluate
$$\int_{C} F_{RR} =  \sum_j q_j\; \#(D_j \cap C) = \sum_j q_j Q_j^C.
$$

By the usual IIA/M-theory duality, it follows that M theory, compactified on circle fibered $X_d$,
is dual to IIA on the base of the circle fibration. Moreover the curvature of
the circle fibarion, $F_{RR}$ above, becomes the RR 2-form flux on $X_{d-1}\times{\bf R}$. In the present context,
we are interested in this duality where $d=4$, and we are relating M-theory on Calabi-Yau fourfold $X_4$
to type IIA on Calabi-Yau threefold $X_3$, with flux, fibered over ${\bf R}$,
$$
{\rm M-theory \;on\;} X_4\qquad \Leftrightarrow \qquad {\rm IIA \;on\;}{X_3 \times{\bf R}.}
$$

Before we go on, we must note a few subtleties. Firstly, as explained in \witten , the linear sigma model describes not just the manifold, but also its moduli space.  However, as explained in \ref\WittenQB{E.~Witten,
  ``Phase Transitions In M-Theory And F-Theory,''
  Nucl.\ Phys.\  B {\bf 471}, 195 (1996)
  [arXiv:hep-th/9603150].
}, the non-geometric phases are absent in M-theory. So, the linear sigma models that describe not only the Calabi-Yau fourfolds for M-theory compactifications, but also the moduli space of M-theory are rather special --  or at least, they are not the same ones as conventionally discussed in IIA. Secondly,
when describing fibrations with $X_3$ used as a base, we would like $X_3$ as well to be in a geometric phase for all values of $r_0$. Otherwise, the interpretation of our construction is less obvious. This constrains the types of charges $Q^0$ we can consider. Thirdly, we have not excluded the possibility that, not only RR fluxes, but also their sources would be turned on on $X_{d-1}$. In other words, sometimes, the IIA dual would involve D6 branes\foot{A simple example of this is the simplest case of our construction. Namely, take $d=2$ and consider ${\bf C}^2$, with coordinates $x_1,x_2$ as a fibration over ${\bf C} \times {\bf R}$
where we introduce a D-term constraint $|x_2|^2 - |x_1|^2 = r_0$, and divide by $(\theta_0, x_1,x_2) \sim (\theta_0 + \lambda, x_1 e^{i \lambda}, x_2 e^{-i \lambda}).$  It is easy to see that $F_{RR} = \delta(x_1) dx_1 \wedge d {\bar x_1}$  has flux through the two sphere surrounding the origin of ${\bf C}\times {\bf R}$. In other words, there is a D6 brane sitting at the origin. This is as expected, since the quotient by $\theta_0$ corresponds precisely to viewing ${\bf C}^2$ as Taub-Nut.}. There is a-priori nothing wrong with this, but for our purposes, we are interested in those cases without the explicit D6 branes. It would be nice to have a direct, geometric, classification of linear sigma models and their circle reductions, that satisfy all of these constraints.

Instead of tackling this problem head on, we will use string theory as a guide.
In the next sections, we will construct the linear sigma models for $X_3$ as low energy effective field theories on the D2 brane probes in IIA string theory. We will explain how turning on fluxes and fibering $X_3$ over ${\bf R}$ modifies the linear sigma model. Moreover, we will show that the resulting manifolds can be viewed as a Calabi-Yau fourfold $X_4$ fibered over ${\bf X}_3\times {\bf R}$. It is well known that D-branes probe only geometric phases of the Calabi-Yau manifolds \orbifolds . This will automatically solve the problems one and two above. Moreover, since we only consider those deformations that correspond to turning on RR fluxes through cycles on $X_3$, there are no D6 branes present by construction.

\newsec{String Duality and D-branes on Local Calabi-Yau}

Consider the theory on $N$ D2 branes probing a local toric Calabi-Yau $X_3$, near the place where $X_3$ develops a singularity. Before turning on fluxes or fibering, this turns out to be a well known quiver theory gauge in three dimensions, preserving ${\cal N}=2$ supersymmetry. As explained in the introduction, the theory is the same as the theory on D3 branes probing $X_3$, dimensionally reduced to three dimensions. The latter was studied in numerous places, for example \refs{\GUD,\FrancoRJ,\FrancoSM,\KV}.
After reviewing this, we consider turning on RR fluxes through the vanishing cycles of $X_3$. We will show that these generate Chern-Simons terms for the 3d gauge theory. To preserve ${\cal N}=2$ supersymmetry on the branes,
than turning on fluxes has to be accompanied with fibering the Calabi-Yau threefold $X_3$ over the one spatial direction transverse to the branes. This corresponds to turning on ${\cal N}=2$ Chern-Simons terms on the D2-branes. These are precisely the three-fold fibrations that we showed are dual to M-theory on Calabi-Yau fourfolds $X_4$! In particular, IIA/M-theory duality implies that, at strong coupling the D2 brane gauge theory goes over to theory on $N$ M2 branes probing $X_4$. Moreover, at strong coupling the kinetic terms on the gauge fields vanish, so the natural description of this is a quiver Chern-Simons theory with ${\cal N}=2$ supersymmetry.

As a further test, the IIA/M-theory duality implies that the center of mass $U(1)$ gauge field on one D2 brane probing $X_3$ is dual to a compact scalar describing the position of the M2 brane on the M-theory circle. For $N=1$, we can explicitly dualize the center of mass $U(1)$ field on the D2 brane probe of our geometry, and show that the result is a linear sigma model on $X_4$, as expected. Moreover, we can show that the circle is fibered over $X_3$ in a manner consistent with the RR fluxes turned on, thus closing the loop of correspondences.

\subsec{D2 brane gauge theory on $X_3$ }

Before turning on fluxes, or fibering, the local Calabi-Yau manifold $X_3$ preserves $8$ supercharges. The $N$ D2 branes probing this break half of the supersymmetries,
preserving four supercharges, or ${\cal N}=2$ supersymmetry in three dimensions.  The theory on the brane probe is a ${\cal N}=2$ quiver gauge theory with superpotential. The data of the theory can be
decoded from the internal geometry. In particular this means that, at the classical level, the quiver for $N$ D2 brane probes is the same as the quiver for $N$ D0 branes or $N$ D3 branes probing the same geometry -- it is only the dimension of the D-branes that changes. The fact that the kinematics of the theories is the same, is a consequence of T-duality in the non-compact dimensions. This fact was also used recently in \Shamit, in a different context.

When $X_3$ is smooth, the theory on $N$ D2 brane probes, at low enough energies, becomes the same as the theory on $N$ D2 branes probing flat space -- so it is the maximally supersymmetric $U(N)$ gauge theory in three dimensions.
At a singularity, however, the D2 branes split into a collection of fractional branes, carrying charges of D6, D4 and D2 branes wrapping wrapping $4$, $2$ and $0$ dimensional cycles shrinking at the singularity (see for example, \ref\DiaconescuBR{
  D.~E.~Diaconescu, M.~R.~Douglas and J.~Gomis,
  ``Fractional branes and wrapped branes,''
  JHEP {\bf 9802}, 013 (1998)
  [arXiv:hep-th/9712230].
}\hv\GUD) . For each such fractional brane on a vanishing cycle $\Delta_{\alpha}$, we get a node of the quiver. If the D2 brane splits into a collection of $n$ cycles
\eqn\whole{
\sum_{\alpha=1}^n d_{\alpha}\; \Delta_{\alpha}
}
we get a
$$
U(d_1 N) \times \ldots \times U(d_n N)
$$
quiver theory on the branes at the singularity.
Moreover, there are $n_{\alpha\beta}$ arrows from the node $\alpha$ to node $\beta$ corresponding to bifundamental fields $\Phi^a_{\alpha \beta}$,
$$
\Phi^a_{\alpha , \beta} \;: \;\alpha \; \longrightarrow \;\beta,
$$
in
$(d_{\alpha} N, {\overline{d_{\beta} N}})$ representation, with $a = 1, \ldots , n_{\alpha\beta}$ (see next section for examples).
There is a gauge invariant superpotential
$$W(\Phi),
$$
which can only depend on the complex structure moduli of $X_3$.
Finally, the Kahler moduli of $X_3$ enter the D-terms of the theory. In particular, the
Fayet-Iliopoulos parameters $r_{\alpha}$ for the $U(1)$ factors of the gauge group depend on them, as
$$
\sum_{\alpha=1}^n r_{\alpha} \, {\rm Tr} D_{\alpha}.
$$

Each vanishing cycle carries definite D-brane charge.
We can describe the charge of branes wrapping $\Delta_{\alpha}$
in terms of its homology cycles in $H_{even}(X_3)$ as
$$
[\Delta_{\alpha}] = \sum \, Q_6^{a} \; [D_a] + \, Q_4^{j} \; [C_j] +  \,Q_2\;[pt]
$$
where $Q_6$ denotes the D6 brane charge and $D$ the 4-cycle class the D6 branes wrap,
$Q_4$ the D4 brane charge, and $C$ a 2-cycle class, and $Q_2$ the D2 brane charge.
The charges are such that $\sum_{\alpha} d_{\alpha} \Delta_{\alpha}$ in \whole\ has the charge of one D2 brane, and nothing else. Note that some of the brane charges appearing must be negative, since $d_\alpha$ are all non-negative. At the singularity the appropriately chosen combinations of D-branes and anti D-branes can be mutually supersymmetric.
This is precisely what allows the D2 brane to split into fractional branes in the first place.

\subsec{D2 branes in IIA on $X_3$ with flux}

Consider turning on RR 2-form on $X_3$. We can write
\eqn\fluxagain{
F_{RR} = \sum_I q_I\, \omega^{(2)}_I
}
where $q_I$ are integers and $\omega^{(2)}_I$ are two-forms with $\delta$-function support on divisors $D_I$ in $X_3$ ($F_{RR}$ is given as a form here, not a class on $X_3$, so the divisors $D_I$ need not all be compact). We would like to know what effect this has in the D2 brane gauge theory at the singularity.

Consider $N$ D4 branes wrapping a vanishing 2-cycle $C$ in $X_3$.
The RR field enters the theory on the branes via a Wess-Zumino coupling
$$
\int_{C \times R^{2,1}} \;A^{RR} \; {\rm Tr} F \wedge F
$$
Note that, even though $C$ is a shrinking cycle, we are allowed to use the geometric description above, since the coupling is topological, and the metric does not enter.
This coupling gives rise to the Chern-Simons term on the D-brane world volume. Integrating by parts, we can rewrite this as
$$
\int_{C \times R^{2,1}}\; F^{RR}\; {\rm Tr} \,\omega_{CS} = k \int_{R^{2,1}}\; {\rm Tr} \,\omega_{CS}
$$
where $\omega_{CS} = A\wedge dA +{2\over 3} A\wedge A \wedge A$ is the Chern-Simons form, and
$$
k = \int_{C} F_{RR}
$$
To compute $k$, we use the description of $F_{RR}$ in terms of forms with $\delta$ function support in \fluxagain\ to express $k$ in terms of intersection numbers of $C$ with divisors $D_i$,
\eqn\chernkk{
k = \sum_I q_I \; \# ( D_I \cap C).
}
This fractional brane wrapped on $C$ could also carry D2 brane charges. Since these would have entered as a first Chern class of the gauge field on the D4 brane, they cannot affect the Chern-Simons coupling above.

Similarly, if we had a D6 brane wrapping a 4-cycle $D$ the Chern-Simons coupling would be given by
$$
k = \int_{D} F_{RR} \wedge {\rm Tr} \,F.
$$
This reduces to \chernkk, where the role of $C$ is now played by the two-cycle class
dual to ${\rm Tr} F$ inside the divisor $D$. This is where the D4 brane charge is induced.
Note that in the wrapped D6 brane case, the fluxes of RR four-form $G_{RR}$ can contribute to the level as well,
by
$$
\delta k = \int_{D} G_{RR}.
$$
We can similarly represent $G_{RR}$ by a 4-form
$$
G_{RR}  = \sum_A p_A \omega^{(4)}_A,
$$
in terms of 4-forms $\omega^{(4)}_A$ with $\delta$-function support on curves $C_A$, and then
$$\delta k = \sum_A p_A \; \#(C_A \cap D).
$$

In summary, turning on RR fluxes corresponds to turning on Chern-Simons terms.
It is easy to see from the above formulas, that the generic combination of 2- and 4-form fluxes corresponds to choosing generic Chern-Simons couplings ${k}_{\alpha}$ for the nodes of the quiver, subject to
\eqn\condi{
\sum_{\alpha =1}^n  k_{\alpha} = 0.
}
This is a consequence of the fact that the net D6 brane and D4 brane charges of the quiver sum up to zero, by construction.

Now, if we were to turn on just fluxes in the Calabi-Yau, the fluxes would break the supersymmetry on the brane to ${\cal N}=1$. We will return to this case in the last section of the paper.
Suppose however that we want to preserve ${\cal N}=2$ supersymmetry. This is the same amount of supersymmetry preserved by an M2 brane on a Calabi-Yau fourfold.
In this case, the Chern-Simons coupling induced by the flux has to be accompanied with the additional terms, corresponding to
$$
k \int d^4 \theta \;V \, \Sigma(V)
$$
where $V$ is the vector multiplet, and $\Sigma$ is the related linear multiplet\foot{$\Sigma$ is defined in terms of $V$ by $\Sigma = \int_0^1 dt {\bar D}(e^{tV} D e^{-tV})$. In the abelian case, we have simply $\Sigma = {\bar D} D V$.}. The bosonic components of the above are
$$
k \Bigl(\; {\rm Tr} \,\omega_{CS}(A) + 2\, {\rm Tr} \,D \;\sigma\Bigr)
$$
where $\sigma$ is the real scalar field in the vector multiplet, and $D$ is the D-term\foot{The vector multiplet has the expansion $V  = -2 i \theta {\bar \theta} \sigma + 2 \theta \gamma^{\mu} {\bar \theta} + \theta^2 {\bar  \theta}^2 D$.}.

We see from the above that, if we want to preserve ${\cal N}=2$ supersymmetry, in the presence of RR fluxes, (some combination of) Kahler moduli of the Calabi-Yau $X_3$ are no longer parameters in the theory -- but they become variable.
More precisely, the vanishing of the potential in $2+1$ dimensions requires \refs{\MSI,\HZ}
$$
\sigma_{\alpha} \Phi_{\alpha \beta}^a = \Phi_{\alpha \beta}^a \sigma_{\beta}.
$$
To see what this means, consider for simplicity, the case of a single D2 brane, $N=1$ on $X_3$. Then, on the Higgs branch, the above is solved by
\eqn\sg{
\sigma_{\alpha} = r_0 \, {\bf 1}_{d_{\alpha}}, \qquad \alpha =1, \ldots, n
}
for some variable $r_0$. In particular, this means that the Fayet-Iliopoulos term for the gauge group on node $\Delta_{\alpha}$ is given by
\eqn\FI{
r_{\alpha}  = k_{\alpha} r_0
}
Since $\sigma$'s parameterize the positions of the branes along the one ${\bf R}$ direction in ${\bf R}^{3,1}$ transverse to the D2 branes, this means two things: first we should identify $r_0$
with coordinate on ${\bf R}$,
$$
r_0 \qquad \leftrightarrow \qquad {\rm coordinate \; on \;{\bf R}}.
$$
Second, it means that $X_3$ is fibered over ${\bf R}$ in a manner that correlates with the RR fluxes turned on. For $N$ D2 branes we just get the symmetric product version of this, with $\sigma_{\alpha}=\sigma_{\beta}$ describing $N$ branes at different positions on ${\bf R}$.

Now, we showed that the IIA compactification on a local Calabi-Yau
threefold $X_3$ where we turn on RR 2-form fluxes and fiber the manifold over ${\bf R}$
by allowing the Kahler moduli to vary is $dual$ to M-theory on Calabi-Yau fourfold $X_4.$
This $implies$ that the gauge theory describing D2 branes on $X_3$ in this background, flows,
in the IR to theory on M2 branes probing $X_4$.
It is natural to $conjecture$ -- based on string/M-theory duality --  that the theory on $N$ M2 branes probing $X_4$ is the quiver Chern-Simons theory obtained from the above by setting the gauge kinetic terms to zero, since these vanish in the IR on dimensional grounds\foot{Note that we can also relax the condition \condi . This corresponds to turning on RR zero form flux, which couples to the D2 brane charge. This agrees with the proposal recently in \refs{\GTI,\GTII}, for Chern-Simons theories dual to massive IIA backgrounds.}.

\subsec{One D2/M2 brane}

As additional evidence for the proposed duality, consider again the case of a single D2 brane in IIA. It is well known that the $U(1)$ gauge field
$A_0$
on the D2 brane probe is dual, via the scalar-vector duality in three dimensions to the compact
scalar
$
\theta_0
$
related to $A_0$ via
\eqn\dual{ d A_0 = * d \theta_0.
}
This scalar describes the position of the M2 brane on the M-theory circle. Based on the IIA on $X_3$/M theory on $X_4$ duality, we expect that when we turn on ${\cal N}=2$ Chern-Simons terms on the branes,
$\theta_0$ should be fibered over the base $X_3 \times {\bf R}$
in such a way so as to give a Calabi-Yau fourfold $X_4$. Moreover, the circle fibration should have the right first Chern class on $X_3$ to correspond to RR 2-form flux we had started with.
We will see in detail that this is indeed the case in the examples we study in the next section.
For now, we will just sketch how this comes about.

For most singularities, and all the singularities we will study in this paper,
$d_{\alpha} = 1$, for all $\alpha$\foot{For examples
where not all $d_{\alpha}=1$ see \GUD .}.
In this case the theory is abelian, with gauge group
$$
U(1)^n
$$
for a quiver with $n$ nodes.

The moduli space is the space of solutions to minima of F and D-term potentials,
modulo the gauge transformations.
Before turning on fluxes, the moduli space of this theory, the Higgs branch, is the Calabi-Yau manifold $X_3 \times {\bf R}$. As above, the ${\bf R}$ direction is parameterized by one massless combination of $\sigma$'s.
On the Higgs branch, all components of the gauge fields with charged matter become massive.
However, there is one decoupled $U(1)$ gauge field, under which nothing is charged. This corresponds to the diagonal combination
$$
A_0 = {1\over n} \sum_{\alpha} A_{\alpha}.
$$
This component becomes the $U(1)$ gauge field on the D2-brane at a generic point in $X_3$.
Dualizing $A_0$ to the compact scalar $\theta_0$ via \dual , the moduli space of the M2 brane is a direct product
${\bf S}^1 \times X_3\times {\bf R}$.
Moreover, the Fayet-Iliopoulos terms in the gauge theory are identified with the Kahler moduli of $X_3$, and they are independent of where in ${\bf R}$ we are.

Now consider turning on 2- and 4-form fluxes. This generates Chern-Simons terms
$$
\int d^3x  \sum_{\alpha=1}^n\, k_{\alpha} \,A_{\alpha}\wedge d A_{\alpha}
$$
where $\sum_{\alpha} k_{\alpha} = 0$, as we explained. The effect of this on the moduli space is twofold.  On the one hand, turning on fluxes modifies the Fayet-Iliopoulos terms in the vacuum corresponding to fibering the $X_3$ over ${\bf R}$ as in \FI .
The effect on the gauge transformations is as follows. Put, $A_\alpha = A_0 + {\hat A}_{\alpha}$,
where $\sum_{\alpha} {\hat A}_{\alpha}=0$.  Then $A_0$ enters the Chern-Simons terms via the coupling\foot{Note that the other Chern Simons terms vanish on the Higgs branch,
since all the components of the gauge fields under which matter is charged are massive.}
\eqn\CSRR{
\int_{{\bf R}^{2,1}} d ^3x  \, \sum_{\alpha} \, k_{\alpha} \, {\hat A}_{\alpha} \, \wedge\, d A_0.
}
Dualizing $A_0$ to the compact scalar $\theta_0$, replaces $dA_0$ by $d \theta_0$. Presence of \CSRR\ means that $\theta_0$ picks up a logarithmic charge $k_\alpha$ under the gauge group on node $\alpha$.
In particular, while $d\theta_0$ is no longer well defined, the combination
$$
d\theta_0 +i \sum_{\alpha} \, k_{\alpha} \, {\hat A}_{\alpha}
$$
is well defined.

We can view this in two different ways. Firstly, this implies that the moduli space is
Calabi-Yau fourfold $X_4$.
It is
obtained by $not$ dividing by the $U(1)$ gauge transformations generated by
$$
\sum_{\alpha} \, k_{\alpha} \, {\hat A}_{\alpha} ;
$$
these we can use to set $\theta_0$ to zero\foot{When the greatest common denominator of the $k_{\alpha}$'s is not $1$, this leaves a discrete subgroup of the gauge symmetry unbroken, this leaves over a discrete subgroup of the gauge symmetry unbroken. The moduli space is then an orbifold by this symmetry.}. This is related by supersymmetry to the one linear combination of $D$-terms we need not impose: we can use them to solve for $r_0$, the superpartner of $A_0$ and $\theta_0.$

Second, this allows us to view the Calabi-Yau fourfold $X_4$ as a circle fibration over $X_3 \times {\bf R}$, with $\theta_0$ parameterizing the fiber.
We will now argue that we can think of $\sum_{\alpha}k_{\alpha}\, F_{\alpha}$
as the RR flux itself, or more precisely that
\eqn\fluxagain{
[F_{RR}]= \sum_{\alpha}  k_{\alpha}[ D_{\alpha}],
}
where $D_{\alpha}$ is the divisor class in $X_3$ corresponding to the $U(1)$ on the node $\alpha$\foot{$D_{\alpha}$ corresponds to setting to zero a product of linear sigma model variables with coupling with charge $1$ to ${\hat A}_{\alpha}$, and charge zero to the rest.}. To see this \AV\ recall first of all that
on the D2 brane there is a Wess-Zumino coupling
$$
\int_{{\bf R}^{2,1}} d^3 x \;F_{RR} \wedge A_0.
$$
In the presence of the above flux, the action of a D2 brane wrapped on a curve $C$ would get shifted by
$\sum_{\alpha} \#( k_{\alpha}D_{\alpha}\cap C)$ times $\int A_0.$

This is exactly what \CSRR\ implies. Namely, configurations of D2 brane wrapped on a curve $C$ in the linear sigma model carries non-zero vortex charge \witten . In particular, a D2 brane wrapped on $C$
carries vortex charge
$$\int F_{\alpha} = \#(D_{\alpha} \cap C)
$$
for the $U(1)$ on node $\alpha$. Correspondingly,
in the presence of the coupling \CSRR\ the action of a D2 brane wrapped on $C$ gets
modified precisely by the same amount, as it does in the presence of the RR flux \fluxagain , as we claimed.
In the next section, we will work this out explicitly in some examples.

Before we go on, note that the $N=1$ case gives a physical realization to the linear sigma models in the previous section\foot{A-priori, the linear sigma model in question differs from the ones in secion 2 by the presence of superpotentials. This is a technicality. In fact, as we'll see in the next section, the solutions to $F-$term equations can be often rewritten as
minima of an {\it effective} linear sigma model without superpotential, \orbifolds . This fact was used extensively in \feng . In such cases the theories on the D-branes are called toric. There are also non-toric examples, related to the toric ones by Seiberg dualities.}. We will see this in the examples in following sections. More precisely,
as noted in \orbifolds , D-branes tend to see only the geometric phases of the theory. This is just what is needed for our purposes. Namely, as argued in \ref\WittenQB{E.~Witten,
  ``Phase Transitions In M-Theory And F-Theory,''
  Nucl.\ Phys.\  B {\bf 471}, 195 (1996)
  [arXiv:hep-th/9603150].
}, the non-geometric phases are absent in M-theory as well.

In the next sections we present some examples of the theories we discussed above.
All of the examples below have appeared in the literature, we are just giving them a new interpretation, in terms of the IIA/M-theory duality.

\newsec{Example 1: Conifold and the ABJM theory}

The Calabi-Yau $X_3$ is the total space
of $O(-1)\oplus O(-1)$ bundle over ${\bf P}^1$. The ${\bf P}^1$ corresponds to the one shrinking 2-cycle of this geometry.
There are two fractional branes,
$$
\Delta_1, \qquad \Delta_2,
$$
whose D-brane charges are given by
\eqn\chone{
[\Delta_1] = [pt] - [{\bf P}^1], \qquad  [\Delta_2] = [{\bf P}^1].
}
In other words, $\Delta_2$ corresponds to a D4 brane wrapping the ${\bf P}^1$,
and $\Delta_1$ to the anti-D4 brane bound to one unit of $D2$ brane. The latter can be thought of as turning on $\int_{{\bf P}^1} F = -1$ on the anti-D4 brane.
Thus, one D2 brane probing $X_3$ corresponds to
$$
 [\Delta_1] + [\Delta_2].
$$
The quiver is given by $U(N) \times U(N)$, ${\cal N}=2$ gauge theory in 3 dimensions, with two pairs of bifundamentals $A_{1,2}$ in $(N, {\bar N})$, and anti-bifundamentals $B_{1,2}$ in $({\bar N}, N)$, and superpotential
$$
W =  \lambda \;  (Tr A_1B_1 A_2B_2 - Tr A_1 B_2 A_2 B_1).
$$
The Higgs branch of the theory describes $N$ D2 branes probing $X_3\times {\bf R}$. Namely, for ${N}=1$, the superpotential vanishes. The potential for the $\sigma_{1,2}$ vanishes by setting
\eqn\vs{
\sigma_1=\sigma_2 =r_0,
}
where we view $r_0$ as a coordinate on ${\bf R}$ inside ${\bf R}^{3,1}$. The D-term potential vanishes for
\eqn\dtc{
|A_1|^2 + |A_2|^2 - |B_1|^2 - |B_2|^2 = r_1,
}
where $r_1$ is the Fayet-Iliopoulos term for the first $U(1)$. For the second $U(1)$ the charges are opposite, and $r_2=-r_1$.  Dividing by the gauge group, the moduli space is the conifold, $X_3$ times ${\bf R}$. While the off-diagonal gauge fields are Higgsed, the diagonal $U(1)$ survives, since nothing is charged under it. It is identified with the center of mass $U(1)$ gauge field on the D2 brane at a generic point in the moduli space. For general $N$, the Higgs branch is the symmetric product $Sym^N(X_3 \times {\bf R})$
describing $N$ identical D2 branes moving about on $X_3 \times {\bf R}$.

Consider now deforming the theory by ${\cal N}=2$ Chern-Simons terms
for the two gauge groups,
\eqn\csc{
k\; \int d^4 \theta \Bigl( {\rm Tr} \; V_1 \Sigma(V_1) - {\rm Tr} \;V_2 \, \Sigma(V_2)\Bigr)
}
where $V_{1,2}$ are the vector multiplets corresponding to the two nodes. This is precisely the theory studied in \ABJM\foot{More precisely, it is the same theory provided we tune $\lambda=1/k$, when the theory has an enhanced supersymmetry.}.
Turning on ${\cal N}=2$ Chern-Simons terms in the D-brane theory means that we are deforming the bulk IIA theory by turning on RR 2-form flux in through the ${\bf P}^1$, and fibering $X_3$ over ${\bf R}$. To see this, consider again the moduli space of the theory for $N=1$. In this case, the Chern-Simons interaction makes Fayet-Iliopoulos parameters dynamical, with
$$
r_1 = k \, r_0 = -r_2
$$
where $r_0$ is the vev of the $\sigma$ fields in \vs .
We still have to divide by the gauge group. Dualizing the center of mass gauge field
to a compact scalar $\theta_0$, the latter picks up a logarithmic charge $k$ under the off diagonal $U(1)$. Thus the gauge symmetry
acts as
\eqn\gc{
{\eqalign{
\theta_0 &\rightarrow \theta_0 + k \lambda\cr
A_{1,2} & \rightarrow A_{1,2} \; e^{i \lambda}\cr
B_{1,2} & \rightarrow B_{1,2} \; e^{-i \lambda}
}}
}
Solving the D-term constraint for $r_0$, and using \gc\ to set $\theta_0$ to zero, the moduli space is
a copy of ${\bf C}^4$, parameterized by $A_{1,2}$, and $B_{1,2}$. We still have to divide by the discrete gauge transformations that leave $\theta_0$ invariant, corresponding to $\lambda =2\pi /k$. Correspondingly, the moduli space is the fourfold
$$
X_4 = {\bf C}^4/{\bf Z}_k,
$$
as explained in \ABJM .
On the other hand, this allows us to view $X_4$ as an $S^1$ fibration over $X_3 \times {\bf R}$.
Namely, projecting to the the base, by forgetting $\theta_0$, and fixing a point $r_0$ in ${\bf R}$, we get a copy of $X_3$. Now consider the first Chern class of the fibration. While $d\theta_0$ itself is not well defined, for example
$$
d \theta_0 + i k \; dB_1/B_1
$$
is well defined over $X_3 \times {\bf R}$. Hence, we can identify
$$
A_{RR} = k \; d B_1/B_1.
$$
The corresponding curvature $F_{RR} = d A_{RR}$ is
$$F_{RR} =  k\, \omega_{B_1}
$$
with
$\omega_{B_1} = i  \delta^2(B_1)\,  dB_1 \wedge d {\overline B_1}.
$
The cohomology class of $F_{RR}$ is the same as $k$ copies of divisor
$D_{B_1}$ corresponding to setting $B_1=0$,
\eqn\class{
[F_{RR}] = k [D_{B_1}].
}
As we explained in the previous section, the RR fluxes turn on Chern-Simons terms on the brane. Given the D-brane charges \chone , we have
$$
k_2 =  \int_{\bf P^1} F_{RR}= - k_1
$$
Consistency requires that the Chern-Simons terms induced by the flux agree with
\csc .
As we explained, given \class\ we have
$$
\int_{\bf P^1} F_{RR} = k\; \#(D_{B_1} \cap {\bf P}^1).
$$
We can read off the intersection number from \dtc . Namely, \dtc\ is the D-term constraint corresponding to the class of the ${\bf P}^1$, so the intersection of the divisor $D_{B_1}$ with the ${\bf P}^1$ is just the charge of $B_1$ under the corresponding $U(1)$. This is $-1$, so
$$
\#(D_{B_1} \cap {\bf P}^1) = -1,
$$
giving $k_1=k = -k_2$, as we had expected.

\newsec{Example 2: ${\bf C^2/Z_2}\times {\bf C}$ singularity}

The Calabi-Yau
$$
X_3 = {\bf C^2/Z_2}\times {\bf C}
$$
is an orbifold, whose resolution is an $O(-2) \oplus O$ bundle over a ${\bf P}^1$. Correspondingly, there is one vanishing two-cycle, the ${\bf P}^1$ in the base.
The theory on the D-branes on $X_3$ was discovered in \DM . There are two fractional branes,
$$\Delta_1, \qquad \Delta_2,
$$
with
$$
[\Delta_1] = [pt] - [{\bf P}^1] \qquad  [\Delta_2] = [{\bf P}^1].
$$
The quiver is given by $U(N) \times U(N)$ gauge theory with a pair of bifundamental chiral multiplets
$A_1,B_1$ in $(N, \bar{N})$, anti-bifundamentals $A_{2}, B_2$, in $({\bar N}, N)$ and an additional pair of adjoints $\Phi_{1,2}$ of the two groups. The
superpotential\foot{In fact the conifold theory in the previous subsection above was obtained as a deformation of the orbifold theory. In the gauge theory, the deformation corresponds to adding a mass term ${m \over 2}  ({\rm Tr} \Phi_1^2 - {\rm Tr} \Phi_2^2)$, with $m=1/\lambda$.} is
$$
W = {\rm Tr} \Phi_1 (A_1 A_2 - B_1B_2)  -{\rm  Tr} \Phi_2 (A_2 A_1 - B_2 B_1).
$$
This theory was studied in the present context in \HZ . For a single D2 brane on $X_3$, the moduli space is $X_3 \times {\bf R}$. The ${\bf R}$ direction is parameterized by $\sigma_1=r_0 =\sigma_2$, as before. $X_3$ emerges by setting the $F-$ and the $D-$terms to zero and dividing by the gauge group. Setting the F-terms to zero we have $\Phi_1=\Phi_2$, which parameterize a copy of ${\bf C}$. In addition, we have
\eqn\fterm{
A_1 A_2 = B_1B_2.
}
The D-term for the first $U(1)$ gives
$$
|A_1|^2 + |B_1|^2 - |A_2|^2 - |B_2|^2 = r_1,
$$
and similarly for the second $U(1)$ with $r_1=-r_2$, and signs of all the charges reversed.
It is useful to rewrite the F-term constraints as D-term constraints, so that we get a linear sigma
model without superpotential. We can solve \fterm\ introducing four new variables, and putting
$$A_1 = x_1 x_0 , \qquad A_2 = x_2 x_3, \qquad B_1 = x_1 x_0 , \qquad B_2 = x_2 x_3.
$$
There is a redundancy inherent in this, which we can remove by simultaneously introducing a $new$ $U(1)$ gauge field
$U(1)_{aux}
$
under which $A$'s and $B$'s are neutral, and $x_{1,2}$ have charge $+1$, and $x_{0,3}$ charge $-1$. This introduces a new D-term constraint
\eqn\ax{
|x_1|^2 + |x_2|^2 - |x_0|^2 - |x_3|^2 = 0.
}
The FI term for this auxiliary $U(1)$ has to be set to zero for this to be equivalent to the solutions of F-term equations. The fields $x_i$ also carry charges under the original gauge fields, since the $A$'s and the $B'$s did. The original D-terms translate to
\eqn\or{
|x_3|^2 -|x_0|^2 = r_1.
}
For $r_1\geq 0,$ we can use \or\ and the corresponding gauge symmetry to solve for $x_3,$
so the moduli space is simply
\eqn\axa{
|x_1|^2 + |x_2|^2 - 2 |x_0|^2 = r_1.
}
modulo $U(1)$. This, is a copy of (the resolution of) ${\bf C}^2/Z_2$, and together
with a copy of ${\bf C}$ parameterized by the adjoints, it gives $X_3$.

As an aside, note that the orbifold phase is absent. This would have corresponded to taking $r_1<0$. However, we are not allowed to do that, since we can no longer solve \or\ in the manner we did before. Instead, what we need to do is exchange the roles of $x_0$ and $x_3$, and then we recover a geometric phase again. As we discussed before, this is just what is needed for compactification of M-theory, since there the non-geometric phases have to be absent \witten .

Consider now turning on ${\cal N}=2$ Chern-Simons couplings for the two gauge groups
of the quiver
$$
k\; \int d^4 \theta \Bigl( {\rm Tr} \; V_1 \Sigma(V_1) - {\rm Tr} \;V_2 \, \Sigma(V_2)\Bigr)
$$
The flux has two effects. The FI parameter of \axa\ becomes dynamical,
$$
r_1= k \, r_0 = -r_2
$$
and, dualizing the center of mass gauge field $A_0 ={1\over 2}( A_1 +A_2)$ to a dual scalar $\theta_0$,
the corresponding gauge transformation becomes
$$\theta_0 \rightarrow \theta_0 + k \lambda,\qquad  x_3 \rightarrow x_3 e^{i \lambda}, \qquad x_0 \rightarrow x_0 e^{-i \lambda}x_0
$$
with $x_{1,2}$ invariant. This allows us to solve for $r_0,\theta_0$ an drop these equations. Their only remnant is a discrete gauge symmetry that takes
$$
x_3, x_0 \rightarrow e^{2 \pi i /k} x_3, e^{-2 \pi i/k}  x_0
$$
The D-term and gauge symmetry corresponding to the auxiliary $U(1)_{aux}$ in \ax\ are unaffected. The moduli space is a Calabi-Yau fourfold $X_4$ which is ${\bf Z}_k$ orbifold of the conifold times a copy of ${\bf C}$ \refs{\HZ,\HVZ}.

Now, let us view $X_4$ as an $S^1$ fibration over $X_3 \times {\bf R}$. We will show that the first Chern class of the fibration is just what is needed for the RR flux to be the origin of the Chern-Simons terms. Namely, it is easy to see that
$$
d\theta_0 + i k( d x_2/x_2 + dx_0/x_0)
$$
is invariant under the $U(1)^2 \times U(1)_{aux}$ gauge group. Thus,
compactifying M-theory on $X_4$ and interpreting $\theta_0$ as the M-theory circle, we get
IIA on ${\bf X_3} \times {\bf R}$ with RR flux turned on
corresponding to
$$
[F_{RR}] =  k [D_{2}] + k [D_{{0}}]
$$
where $D_{i}$ is a divisor corresponding to setting $x_i=0$. Now,
the Chern-Simons level should be given by
$$
k_2 = -k_1 = \int_{{\bf P}^1} F_{RR}.
$$
We have
$$
\int_{{\bf P}^1} F_{RR} = k\; \#((D_{2}+D_0) \cap {\bf P}^1)
$$
and we would like to compute what this is on $X_3$. Since the D-term constraint corresponding to the class of the ${\bf P}^1$ is \axa\ the intersection numbers are simply the charges of $x_2$, $x_0$ under it. This gives
$$
\#(D_{2}\cap {\bf P}^1) = 1, \qquad \#(D_{0}\cap {\bf P}^1) = -2
$$
we see that we recover
$$
k_1 = k = -k_2,
$$
as expected.

\newsec{Example 3: The Suspended Pinch Point}

In this case, the $X_3$ geometry has two vanishing ${\bf P^1}$'s, ${\bf P}^1_1$ with normal bundle $O \oplus O(-2)$ bundle, and ${\bf P}^1_2$  with normal bundle $O(-1) \oplus O(-1)$. There are now three fractional branes.
\eqn\chtwo{
[\Delta_1] = [{\bf P}^1_{1}], \qquad [\Delta_2] = [{\bf P}^1_{2}], \qquad [\Delta_3] =   - [{\bf P}^1_{1}]- [{\bf P}^1_{2}]+[{\rm pt}]
}
For $N$ D2 branes probing $X_3$ we get a
$$U(N)\times U(N) \times
U(N)
$$
quiver theory with bifundamental matter as in the figure 1.
Since the brane in class $\Delta_1$ is free to move in the $O$ direction, (the normal bundle has one holomorphic section) there is one adjoint chiral multiplet $\Phi$ corresponding to that. The superpotential is
$$ W = Tr \Phi( A_1 A_2 - C_1 C_2) + Tr B_1 B_2 C_2 C_1 - Tr B_2 B_1 A_2 A_1.
$$
\bigskip
\centerline{\epsfxsize 2.5truein\epsfbox{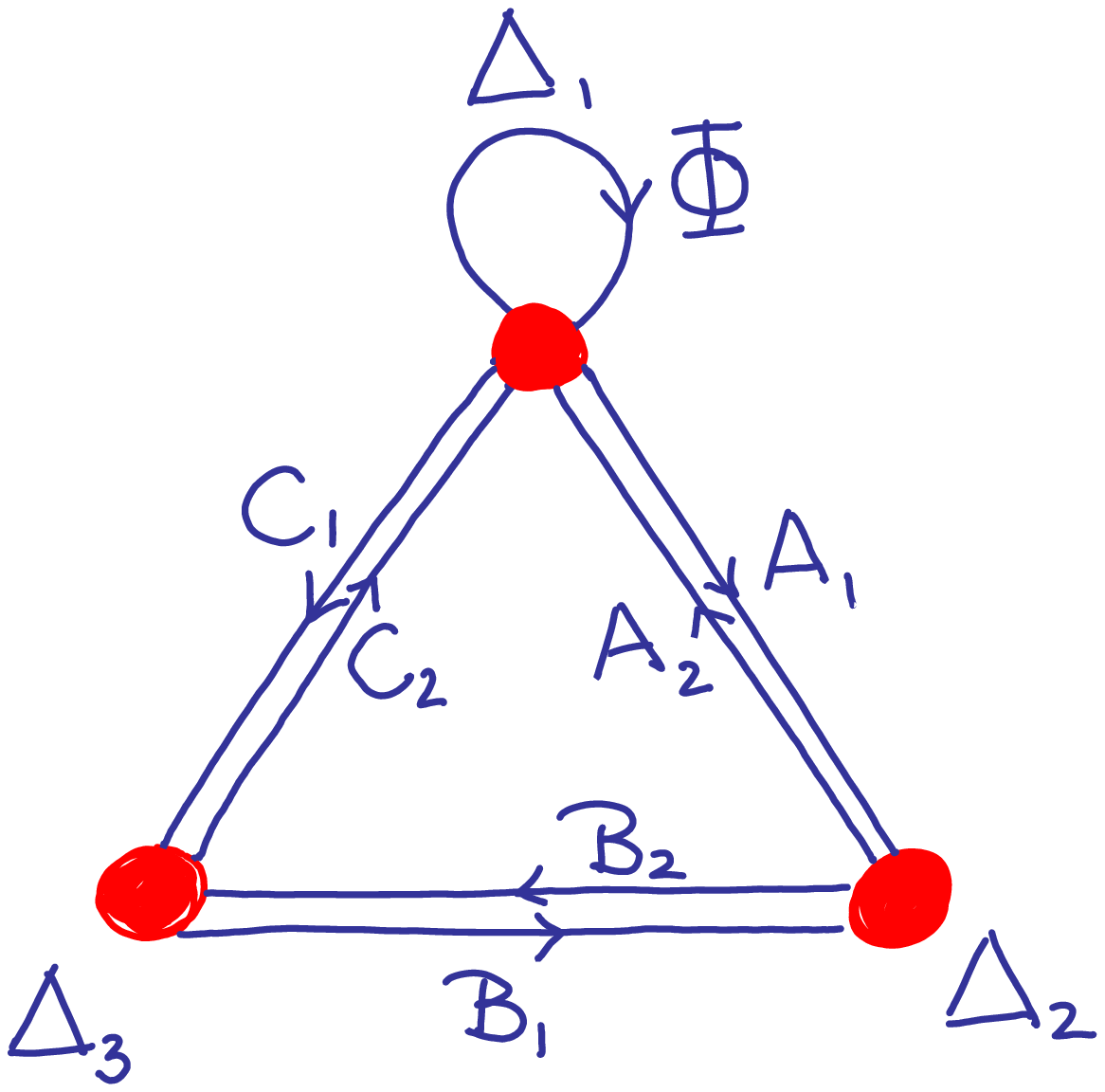}}
\noindent{\ninepoint \baselineskip=2pt {\bf Fig. 1.}{ Quiver corresponding to the $N$ D2 branes on the Suspended Pinch Point singularity. } }
\bigskip
Consider the moduli space for one D2 brane on $X_3$. As usually, there is one direction ${\bf R}$ parameterized by $\sigma_{\alpha}=r_0$, for all $\alpha$.
The F-term equations set
$$
A_1 A_2 = C_1 C_2, \qquad \Phi = B_1B_2
$$
We can write these as D-term equations by putting
$$
A_1 = a_1b_1, \qquad A_2 = a_2b_2, \qquad C_1 = a_1 b_2, \qquad C_2 = a_2 b_1,
$$
and introducing an additional $U(1)_{aux}$ gauge symmetry, under which $a_{1,2}$ have charge $+1$ and $b_{1,2}$ have charge $-1$. All in all, we have $U(1)^4$ gauge invariance, and four D-term equations
\eqn\spp{\eqalign{
&|a_1|^2-|a_2|^2 = r_1\cr
&|a_2|^2-|b_1|^2-|B_1|^2 +|B_2|^2 = r_2\cr
&|a_2|^2-|b_2|^2+|B_1|^2 -|B_2|^2 = r_3\cr
&-|a_1|^2-|a_2|^2+|b_1|^2 +|b_2|^2 = 0
}
}
where $r_1+r_2+r_3=0$. It is easy to show, solving the D-term equations and dividing by the gauge group, that the moduli space is precisely $X_3$ with $r_1$ corresponding to the size of ${\bf P}^1_1$
and $r_2$ to ${\bf P}^1_2$. Namely, using the first equation, and the corresponding gauge symmetry to solve for $a_1,$ the moduli space can be written simply as
\eqn\spptwo{\eqalign{
&|b_1|^2+|b_2|^2-|a_2|^2 = r_1\cr
&|a_2|^2-|b_1|^2-|B_1|^2 +|B_2|^2 = r_2\cr
}}
where the first equation now corresponds to the cohomology class of the ${\bf P}^1_1$
and the second to ${\bf P}^1_1$.

Consider now turning on ${\cal N}=2$ Chern-Simons couplings. Given the D-brane charges \chtwo\  the Chern-Simons couplings should be related to fluxes as
$$
k_1 = \int_{\bf P^1_1} F_{RR},\qquad  k_2 = \int_{\bf P^1_2} F_{RR}, \qquad k_3 = - \int_{\bf P^1_1} F_{RR} - \int_{\bf P^1_2} F_{RR}.
$$
\subsec{ Case $(k_1,k_2,k_3) = (k, -k,0)$}

The Fayet-Iliopoulos terms $r_1$, $r_2$ vary over ${\bf R}$ parameterized by $r_0$ as
$$
r_1 \sim r_0, \qquad r_2 \sim  -r_0,
$$
while $r_3$ is a constant.  We can view $X_4$ obtained by just dropping the first two equations in \spp\ and the corresponding $U(1)$'s, and dividing by a $Z_k$ action that sends $a_1, a_2$ to $e^{2 \pi i/k} a_1, e^{-2 \pi i/k} a_2$. The moduli space is the ${\bf Z_k}$ quotient of the fourfold $D_3$ given in \refs{\HZ,\HVZ}.

On the other hand, we can view  $X_4$ as a circle fibration over $X_3 \times {\bf R}$ by dualizing the center of mass $U(1)$ to a compact scalar field $\theta_0.$
Then, $\theta_0$ is invariant under the third and fourth $U(1)$ in \spp, and transforms under the first two with logarithmic charge $k$, and $-k$, respectively.  We can pick, for example
$$
d \theta_0 - k (da_1/a_1 + d b_1/b_1)
$$
as the invariant one form on $X_3 \times {\bf R}$.  This implies
$$
[F_{RR}]  = k [D_{a_1} + D_{b_1}]
$$
where $D_{a_1}$, $D_{b_1}$ are the divisors corresponding to setting $a_1$, and $b_1$ to zero.
From \spp, and the charges of $a_1$, $b_1$ we can read off that
$$
\#(D_{a_1} \cap {\bf P}^1_1) = 0, \qquad \#(D_{a_1} \cap {\bf P}^1_2) = 0, \qquad
\#(D_{b_1} \cap {\bf P}^1_1) = 1, \qquad \#(D_{b_1} \cap {\bf P}^1_2) = -1
$$
so that
$$
\int_{{\bf P}^1_1} F_{RR} = k = - \int_{{\bf P}^1_2} F_{RR},
$$
as expected.
\subsec{The General Case $(k_1,k_2,k_3) = (k, m,-k-m)$}

The Fayet-Iliopoulos terms $r_1$, $r_3$ vary over ${\bf R}$ parameterized by $r_0$ as
$$
r_1 \sim m r_0, \qquad r_2 \sim  k r_0 \qquad r_3 \sim -(k +m) r_0,
$$
while $r_2$ is a constant.  We can view $X_4$ obtained by dropping the first three equations
and replacing them by one linear combination of them which is independent of $r_0$, say $k$ times the
first equation minus $m$ times the second. In addition, we divide by the corresponding $U(1)$ gauge symmetry. The result is a Calabi-Yau fourfold $X_4$.

Now, let's view  $X_4$ as a circle fibration over $X_3 \times {\bf R}$. We are want to show that the corresponding circle fibration has the correct Chern class.
The compact scalar $\theta_0$ transforms with logarithmic charges $k$, $m$ and $-k-m$ under the first three $U(1)$'s in \spp\ and not at all under the last one.
The invariant one form is
$$
d \theta_0 - (k\; da_1/a_1 -m\;  d b_1/b_1+(k+m) \; d b_2/b_2)
$$
From this it follows that
$$[F_{RR}] = k [D_{a_1}] - m[D_{b_1}] + (k+m) [D_{b_2}].
$$
We know the intersection numbers of $D_{a_1,b_1}$ already. We have in addition
$$
\#(D_{b_2} \cap {\bf P}^1_1) = 1, \qquad  \#(D_{b_2} \cap {\bf P}^1_2) =0
$$
Adding this all up, we find
so that
$$
\int_{{\bf P}^1_1} F_{RR} = k, \qquad  \int_{{\bf P}^1_2} F_{RR} = m,
$$
as expected, given the D-brane charges \chtwo .

\newsec{Example 4: ${\bf C^3/Z_3}$}

In this case, the resolution of the singularity is $O(-3)\rightarrow {\bf P}^2$, so there is both a shrinking four cycle, the ${\bf P}^2$ and a shrinking 2-cycle, the ${\bf P}^1$ inside the ${\bf P^2}$. There are three vanishing cycles $\Delta_{1,2,3}$.
The D-brane charges of vanishing cycles are \GUD\
$$
\Delta_1 = -2 [{\bf P}^2] + [{\bf P}^1] - {1\over 2} [{\rm pt}], \qquad \Delta_2 = [{\bf P}^2] \qquad \Delta_3 = [{\bf P}^2] - [{\bf P}^1] - {1\over 2} [{\rm pt}],
$$
The gauge group on $N$ D2 branes is
$$U(N) \times U(N) \times U(N),
$$
with nine chiral multiplets connecting them, as in the figure 2.
\bigskip
\centerline{\epsfxsize 2.5truein\epsfbox{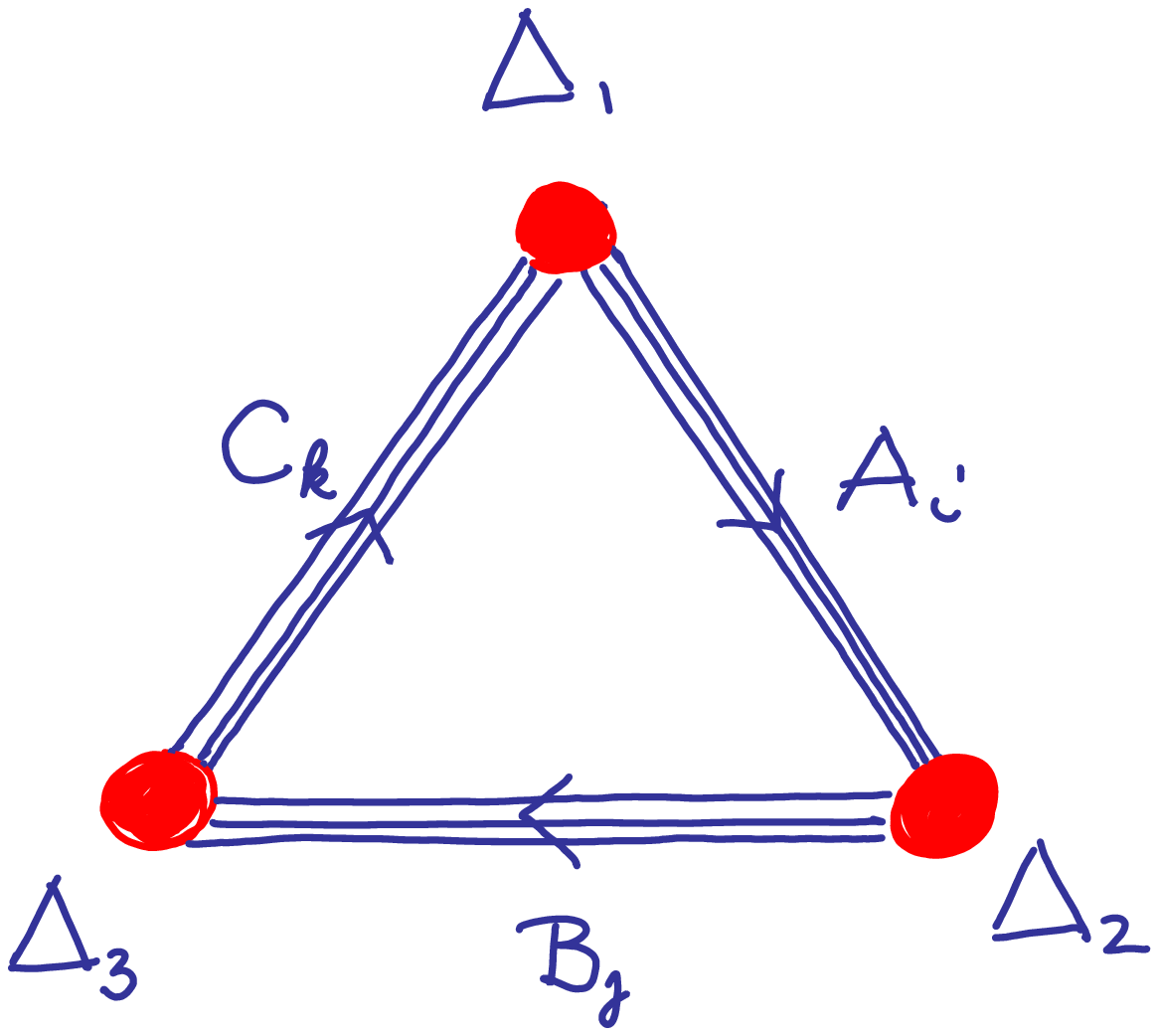}}
\noindent{\ninepoint \baselineskip=2pt {\bf Fig. 2.}{ Quiver corresponding to the $N$ D2 branes on ${\bf C}^3/{\bf Z}_3$. There are three bifundamentals connecting each pair nodes. } }
\bigskip
%
In addition, there is a superpotential
$$W = \sum_{i,j,k} \epsilon_{ijk} {\rm Tr} A_i B_j C_k
$$
Consider now the moduli space for a single D2 brane probe. This problem was solved explicitly in \orbifolds . As in the previous examples, we can redefine variables to be able to rewrite the F-term equations as D-term equations. This example was worked out in detail in \orbifolds . At the end of the day, we have an effective linear sigma model with
six fields $p_i$, $i=0, 1, \ldots, 5$, and $U(1)^4$ gauge symmetry
\eqn\ptwo{\eqalign{
&|p_1|^2+|p_2|^2+|p_3|^2 - |p_0|^2-|p_4|^2-|p_5|^2 = 0\cr
-&|p_4|^2+|p_5|^2 = r_1\cr
&|p_0|^2-|p_5|^2 = r_2\cr
-&|p_0|^2+|p_4|^2 = r_3\cr
}}
where $r_i$ is the FI parameter of the $i$'th node, and $r_1+r_2+r_3=0$.
For any choice of FI parameters, the moduli space is always $X_3=O(-3) \rightarrow {\bf P}^2$, i.e. the orbifold phase is absent. For example, for $r_2<0<r_3$, we can use the last two equations above to "solve" for $p_{4,5}$ in terms of $p_0$, and the moduli space is simply the solutions to
$$
|p_1|^2+|p_2|^2+|p_3|^2 - 3|p_0|^2 = r_3-r_2
$$
which is $X_3$ with Kahler class $r_3-r_2$.

Consider now turning on Chern-Simons couplings for the branes on the three nodes
with
$$
(k_1,k_2,k_3) = (k,- k-m, m).
$$
It is easy to work out what the corresponding Calabi-Yau fourfold $X_4$ is.
This corresponds to letting the FI parameters vary as
$$
r_1 \propto k\;r_0, \qquad r_2 \propto -(m+k)\;r_0, \qquad r_3 \propto m \;r_0,
$$
moreover, the compact scalar $\theta_0$ has charges $k, -(k+m)$ and $-m$ under the three $U(1)$'s
corresponding to the three nodes od the quiver, and is neutral under the auxiliary $U(1)$ (corresponding to first equation in \ptwo).
The invariant one form can be written as %
$$
d \theta_0 + i\, (m \, dp_0/p_0 - k\, d p_5/p_5 +(m-k) dp_1/p_1)
$$
This implies
$$[F_{RR}] = -m [D_0] - (m-k) [D_1] + k [D_5]
$$
From above, we can read off,
$$
\#(D_0 \cap {\bf P}^1) = -3, \qquad \#(D_1 \cap {\bf P}^1) = 1, \qquad \#(D_5 \cap {\bf P}^1) = 0
$$
so
$$\int_{{\bf P}^1} F_{RR} =   2 m +k
$$

Now, consider the contribution of this flux to the Chern-Simons levels. For the three fractional branes $\Delta_1$, $\Delta_2,$ $\Delta_3$, the RR 2-form flux gives:
$$
(k_1,k_2,k_3) = (\int_{\bf P^1} F_{RR}, 0, -\int_{\bf P^1} F_{RR} ) = (2m+k ,0, -2m-k).
$$
Now, recall that we are also free to turn on RR four-form flux on the ${\bf P}^2$ in IIA.
From the charges of the wrapped D-branes, we see that this could shift the levels by
$$
(\delta k_1,\delta k_2,\delta k_3) = (-2 \int_{\bf P^2} G_{RR}, \int_{\bf P^2} G_{RR}, \int_{\bf P^2} G_{RR}),
$$
which is exactly what we need, provided
$$
\int_{{\bf P}^2} G_{RR} = m.
$$

Note that, for $r_0<0$, we get a different assignment of fluxes. As we pass through the singularity at $r_0$, and continue on to negative $r_0$, the fractional branes are permuted by the quantum ${\bf Z_3}$ symmetry, that takes $\Delta_{i}$ to $\Delta_{i-1}$ \ref\DouglasQW{
  M.~R.~Douglas, B.~Fiol and C.~Romelsberger,
  ``The spectrum of BPS branes on a noncompact Calabi-Yau,''
  JHEP {\bf 0509}, 057 (2005)
  [arXiv:hep-th/0003263].
}.  This also acts on the homology in the corresponding way, namely, $([{\bf P}^2], [{\bf P}^1])$ go to
$([{\bf P}^2]- [{\bf P}^1], 3 [{\bf P}^2]- 2[{\bf P}^1]),$  so the Chern-Simons couplings get permuted as well. For this reason, it does not follow that the four-form flux on $X_3$ lifts to $G$-flux on the fourfold.

\subsec{Case $(k_1,k_2,k_3) = (0, -m, m)$}

For example, consider the case where $k=0$.
Note that this does not mean that the RR 2-form flux is turned off entirely on $X_3,$ but just that the its cohomology class vanishes.

In this case, the Calabi-Yau fourfold $X_4$ is an ${\bf Z}_m$ quotient
of an $O(-2)\oplus O(-1)$ bundle over ${\bf P^2}$. Namely, we can use the last equation and the corresponding gauge symmetry \ptwo\ to
"solve" for $r_0$, and $\theta_0.$ The two left-over, independent D-term constraints are the first two. We can solve the latter one for $p_5$, and the manifold becomes
$$
|p_1|^2+|p_2|^2+|p_3|^2 - |p_0|^2-2 |p_4|^2= r_1
$$
modulo the corresponding $U(1)$. This is the $O(-1)\oplus O(-2)$ bundle over ${\bf P}^2$. This agrees with \refs{\HZ, \HVZ}.
More precisely, $X_4$ is a quotient of this
$$
p_0, \, p_4\;\; \rightarrow \;\;e^{2 \pi i /m} p_0, e^{-2 \pi i /m} p_4
$$
which is a discrete gauge symmetry left over from solving for $\theta_0.$ Finally, we claim that the
quiver gauge theory in this case describes M2 branes on $X_4$, with $m$ units of $G$-flux turned on through the ${\bf P}^2$. At large $N$, the theory should be dual to M-theory on $AdS_4 \times Y^{m,2m}({\bf P^2})$ \refs{\MSI, \MSII}.

\newsec{Example 5: Local ${F_0}$ and the cone over $Q^{1,1,1}$}

This case is interesting, since as we will see, it will provide a new Chern-Simons theory
that should be dual to $N$ M2 branes probing a cone over $Q^{1,1,1}$. The later is a very old example of an Einstein-Sasaki manifold. The conformal field theory dual to $AdS_4\times Q^{1,1,1}$ has been sought for a long time. Here we will provide a string theory derivation of the dual CFT. Other proposals have been made in \FHPR\ and studied in \klebanov\ recently.

The geometry of the Calabi-Yau threefold $X_3$ is a line bundle over
$$F_0={\bf P}^1 \times{\bf P}^1.$$
In this case there are four vanishing cycles. There are two different assignments of charges we can make where the superpotential is still toric (in the sense of the footnote 14). We will pick one of these, for simplicity. The D-brane charges are \GUD\
$$\eqalign{
\Delta_1& = [F_0] - [{\bf P}^1_1]-[{\bf P}^1_2]+[{\rm pt}]\cr
\Delta_2& = [F_0] - [{\bf P}^1_1]\cr
\Delta_3& = -[F_0] +2 [{\bf P}^1_1]\cr
\Delta_4& = -[F_0] +[{\bf P}^1_2]\cr}
$$
%
\bigskip
\centerline{\epsfxsize 2.5truein\epsfbox{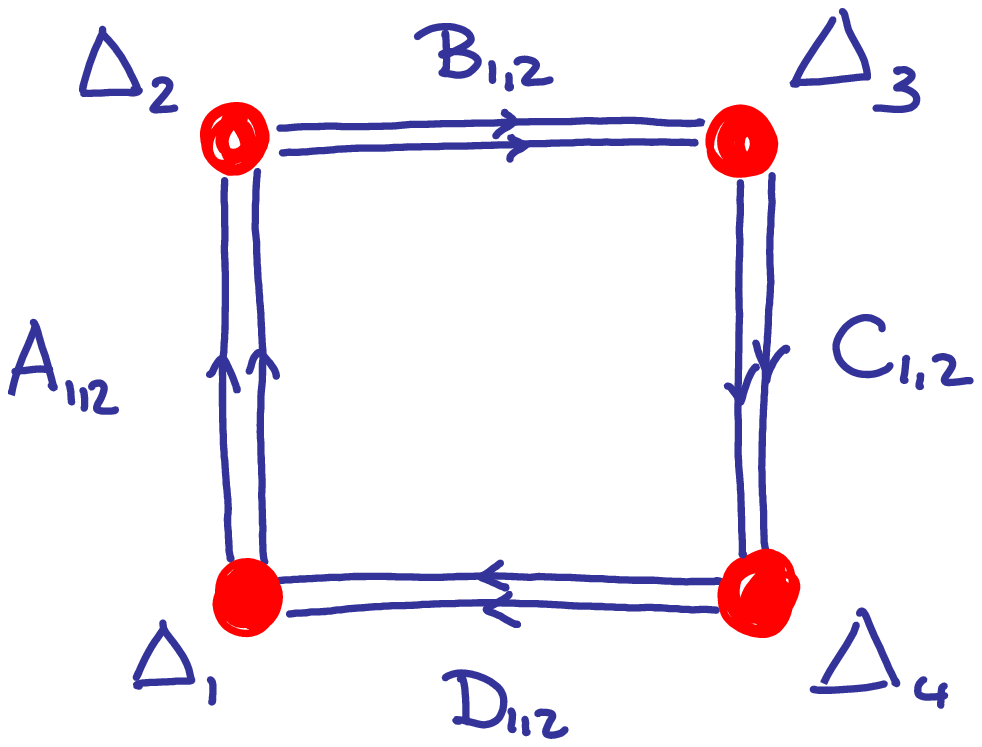}}
\noindent{\ninepoint \baselineskip=2pt {\bf Fig. 3.}{ Quiver corresponding to the $N$ D2 branes on local $F_0$. There are two bifundamentals connecting each pair nodes. } }
\bigskip
The theory on $N$ D2 branes is a
$$ U(N) \times U(N) \times U(N) \times U(N)
$$
quiver theory corresponding to the figure 3,
with superpotential
\eqn\suppof{
W =\epsilon^{ik} \epsilon^{jl} \;{\rm Tr} A_i B_j C_k D_l.
}

Consider the moduli space for $N=1$. Before turning on superpotential, the moduli space is ${\bf R} \times S^1$, parameterized by $\sigma_{\alpha}=r_0$ and $\theta_0$, times $X_3$.
The latter is a space of solutions to F-term equations
$$
A_{1}C_2 = A_2 C_1, \qquad B_1 D_2 = B_2 D_1
$$
and D-term constraints
modulo $U(1)^4$. We can rewrite the F-term constraints as D-term constraints of an auxiliary linear sigma model, by introducing $8$ new variables %
$$
x,\; y,\;z,\;w,\; {\tilde x},\; {\tilde y},\; {\tilde z},\; {\tilde w}
$$
such that
\eqn\redef{\eqalign{
A_1 &= xy, \qquad A_2 = x w, \qquad C_1 = zy, \qquad C_2 = zw\cr
B_1 &= \tx \ty, \qquad B_2 = \tx \tw, \qquad D_1 = \tz \ty, \qquad D_2 = \tz\tw
}
}
The D-term constraints now become
\eqn\ld{\eqalign{
&|x|^2 - |\tz|^2 = r_1\cr
&|\tx|^2 - |x|^2 = r_2\cr
&|z|^2 - |\tx|^2 = r_3\cr
&|\tz|^2 - |z|^2 = r_4\cr
&|x|^2 +|z|^2 - |y|^2 - |w|^2 = 0\cr
&|\tx|^2 +|\tz|^2 - |\ty|^2 - |\tw|^2 = 0\cr
}
}
and the corresponding $U(1)$ actions can be read off from there. The last two D-terms serve to remove the redundancies inherent in \redef . The first four D-terms correspond to the four $U(1)$'s of the quiver. Note that only three of these are independent. There is one constraint on the Fayet-Iliopoulos terms,
$$r_1+r_2+r_3+r_4=0.
$$
For, say $r_{1,2,3}>0$, we can solve for $x, z,\tx$ in terms of $\tz$, and the moduli space is manifestly the line bundle over local $F_0,$ obtained by setting
$$\eqalign{
&|y|^2 + |w|^2 = 2|\tz|^2 + r_1-r_4\cr
&|\ty|^2 + |\tw|^2 = 2|\tz|^2 + r_1+ r_2 \cr
}
$$
modulo the corresponding $U(1)$'s.

Consider now turning on a specific combination of Chern-Simons terms,
$$
(k_1,k_2,k_3,k_4) = (0,k,0,-k).
$$
This corresponds to setting
$$r_2 \sim k r_0, \qquad r_4 \sim -k r_0,
$$
while the others are constant. We can use the second D-term in \ld\ to solve for $r_0,$ and the corresponding gauge symmetry,
\eqn\gaugeoa{\eqalign{
\theta_0 &\rightarrow \theta_0 + k \lambda\cr
x &\rightarrow x e^{-i \lambda}\cr
\tx &\rightarrow \tx e^{i \lambda}\cr}
}
to set $\theta_0$ to zero.
The first and the third D-term express $x$ and $z$ in terms of $\tx$ and $\tz$.
The moduli space is thus
$$\eqalign{
&|y|^2 +|w|^2 = |\tx|^2+|\tz|^2 + r_1+r_3\cr
&|\ty|^2 +|\tw|^2 = |\tx|^2+|\tz|^2\
}
$$
modulo the corresponding $U(1)^2$ action. For $k=1$, and $r_1=0=r_2$, this gives a moduli space which is a cone over $Q^{1,1,1}$!

For general $k$, we still have to divide by the discrete gauge symmetry left over, corresponding to $\lambda = 2 \pi i/k$ in \gaugeoa . A symmetry under which $x$ and $\tx$ transform as $(x, \tx) \rightarrow ( e^{2\pi i/k} x, e^{-2\pi i /k} \tx)$, while the rest of coordinates are neutral is gauge equivalent to
one where $\tx$ and $\tz$ transform as
$$
(\tx,\;\tz) \qquad \rightarrow \qquad( e^{2\pi i/k}\; x, \;e^{-2\pi i /k}\; \tz).
$$
It is easy to see from above, that this corresponds to IIA string theory on $X_3\times {\bf R}$
with
$$
\int_{{\bf  P}^1_1} F_{RR} = k ,\qquad \int_{{\bf  P}^1_2} F_{RR} = k.
$$
where to reproduce the Chern-Simons terms, an additional four-form flux must be turned on,
$$
\int_{F_0} G_{RR} = k .
$$
Note that despite the presence of the 4-form flux in IIA, there need not be any $G$-fluxes turned on in M-theory. This is because, just like in the previous section, there are monodromies acting on the cycles and fluxes as we go from positive to negative $r_0$, under which the 4-form flux transforms. The interpretation of this in the Calabi-Yau four-fold context is not clear to us.

In summary, IIA/M-theory duality predicts that the theory dual to the cone on
$AdS_4\times Q^{1,1,1}/{\bf Z}_k$ should be the $U(N)^4$ Chern-Simons theory corresponding to the quiver
in figure $3$ with superpotential \suppof , and Chern-Simons levels $(0,k,0,-k)$\foot{The quotient $Q^{1,1,1}/Z_k$ is called $Y^{k,k}({\bf P}^1 \times {\bf P}^1)$ in \refs{\MSI, \MSII}.}. Since two of the Chern-Simons levels vanish, it is not obvious the theory has a weak coupling expansion at large $k$. The gauge fields associated to nodes $1$ and $3$ act as constraints, setting the corresponding currents to zero. Since this provides a very simple example of AdS/CFT correspondence, it would be nice to study this theory in detail. The theory may have a dual description in terms of Chern-Simons theories in \refs{\FHPR,\klebanov}, whose direct string theory realization we do not know.

\newsec{$N$ M2 branes on $G_2$ holonomy manifolds}

In this section we conjecture a lagrangian description of $N$ M2 branes probing a local $G_2$ holonomy manifold. For a single M2 brane, the corresponding low energy effective theory was proposed in \AV . We will explain here how this generalizes to arbitrary $N$.
To begin with, consider IIA string theory compactified on a Calabi-Yau threefold $X_3$. The theory has ${\cal N}=2$ supersymmetry in the bulk. Turning on RR 2-form fluxes through the two-cycles of $X_3$ breaks supersymmetry to ${\cal N}=1$ in four dimensions. The IIA string theory on $X_3$ with RR fluxes is dual M-theory on a $G_2$
holonomy manifold. We can get an explicit description of the $G_2$ holonomy manifold \AV\ along the lines of what we did in section two.

Consider again, as we did in section 2, a Calabi-Yau threefold described by a linear sigma model with $N+4$ chiral fields $\phi_i$ and $N+1$ $U(1)$ gauge fields with charges $Q_i^a$, $a=0, \ldots N.$ The Calabi Yau is the space of solutions to D-term equations
$$
\sum_{i} Q_i^q \; |\phi_i|^2 = r_a, \qquad a = 0, \ldots N
$$
modulo the gauge transformations. Consider now turning on RR 2-form flux
$$[F_{RR}] = \sum_{a} k_{a} \;[D_{a}]
$$
where $D_{a}$ is a divisor corresponding to setting to zero a product of variables with charge $1$ under the $a$-th $U(1)$, and zero under the rest.
Type IIA string theory in this background should lift to a $G_2$ holonomy manifold in M-theory. It is easy to write down the corresponding $G_2$ manifold, at least topologically. The fluxes imply that the $G_2$ holonomy manifold is a non-trivial $S^1$ fibration over $X_3$. Borrowing the results of section 2, the $G_2$ manifold is a Hopf-fibration: under the $U(1)^{N+1}$ gauge transformations
that take
$$\phi_i \rightarrow \phi_i \; \exp(i Q_i^a \lambda_a)
$$
the compact scalar $\theta_0$ parameterizing the circle transforms as
$$
\theta_0 \rightarrow \theta_0 + \sum_{a} k_a \lambda_a.
$$
The difference with respect to section 2 is that the Calabi-Yau threefold is no longer fibered over ${\bf R}$ in a non-trivial way. This implies that the Calabi-Yau fourfold geometries of section 2 can also be viewed as $G_2$ holonomy fibrations over the real line. We'll return to this point below.

Consider now $N$ D2 brane probing $X_3$. As we reviewed in section 3, before turning on fluxes, in the limit where $X_3$ develops a singularity, the theory on the branes is an ${\cal N}=2$ quiver gauge theory in three dimensions with superpotential. As we discussed in section 3, turning on RR fluxes turns on Chern-Simons terms on the D-branes. In fact, in the present context, they correspond to turning on ${\cal N}=1$ Chern-Simons terms. This is because the fluxes break half the supersymmetry, so the theory on the branes with fluxes should have ${\cal N}=1$ supersymmetry in three dimensions. The corresponds to deforming the action by
$$
\int_{{\bf R}^{2,1}} d^{3}x \sum_{\alpha} k_{\alpha} \;S_{CS} (A_{\alpha})
$$
where $\alpha$ denotes the nodes on the quiver, and
$$
S_{CS}(A_{\alpha})
 = {\rm Tr} \;\omega_{CS}(A_{\alpha}) - {\rm Tr}\; {\bar \chi}_{\alpha} \chi_{\alpha}
$$
denotes the ${\cal N}=1$ Chern-Simons term in the notation of \Schwarz . The Chern-Simons levels get related to RR 2-form fluxes and the RR 4-form fluxes through the vanishing cycles, just as in the ${\cal N}=2$ case, and moreover satisfy
$$\sum_{\alpha} k_{\alpha} =0.$$

For a single brane probe, the center of mass gauge field on the D2 brane can be dualized to a compact scalar $\theta_0$ parameterizing the position of the M2 brane on the M-theory circle. It is easy to see that, in the presence of ${\cal N}=1$ Chern-Simons terms, the $S^1$ becomes fibered non-trivially over $X_3$ in such a way that the moduli space becomes $G_2 \times \bf{R}$.

For example, take $X_3$ to be the conifold, discussed in section 4. The theory on $N$ D2 branes on $X_3$ is the Klebanov-Witten type quiver theory described there. Now consider turning on RR 2-form flux through the ${\bf P}^1$ of the conifold,
$$ \int_{{\bf P}^1} F_{RR} = k,
$$
but {\it not} fibering $X_3$ over ${\bf R}$, as we did there. This turns on ${\cal N}=1$ Chen-Simons terms with $k_1=k =-k_2$.
Consider now the moduli space, for one D2 brane. The potential for $\sigma$'s vanishes for $\sigma_{1}=r_0 = \sigma_2$ as before, but this no longer affects the FI terms. The D-term potential vanishes for
\eqn\dff{
|A_1|^2+|A_2|^2 - |B_1|^2-|B_2|^2 = r
}
where $r_1=r = -r_2$ is a constant Fayet-Iliopoulos term. We still need to divide by the gauge group. In the presence of the Chern-Simons coupling, the $\theta_0$ scalar dual to the center of mass gauge field picks up a charge, so the gauge transformations act as
$$\eqalign{
\theta_0 &\rightarrow \theta_0 + k \lambda\cr
A_{1,2} &\rightarrow A_{1,2}\; e^{i \lambda}\cr
B_{1,2} &\rightarrow B_{1,2} \; e^{-i \lambda}.}
$$
We can use these to gauge away $\theta_0$, so the moduli space is a 7-manifold described by the locus \dff\ in ${\bf C}^4$ parameterized by $A_{1,2}$, and $B_{1,2}$. For $k=1$, this is precisely the $G_2$ manifold discussed in \ref\AMV{M.~Atiyah, J.~M.~Maldacena and C.~Vafa,
  ``An M-theory flop as a large N duality,''
  J.\ Math.\ Phys.\  {\bf 42}, 3209 (2001)
  [arXiv:hep-th/0011256].
}. For $k\neq 1$, we still have to divide this by the discrete gauge symmetry left over, generated by $\lambda = 2 \pi /k$ so the manifold is a ${\bf Z}_k$ quotient. It is related by the "second" M-theory flop discovered in \ref\AtiyahQF{
  M.~Atiyah and E.~Witten,
  ``M-theory dynamics on a manifold of G(2) holonomy,''
  Adv.\ Theor.\ Math.\ Phys.\  {\bf 6}, 1 (2003)
  [arXiv:hep-th/0107177].
} to the ${\bf Z}_k$ actions considered in \AMV .

Similarly, for all other examples we wrote down in sections 5-8, we get pairs of $G_2$ manifolds
and the corresponding ${\cal N}=1$ quiver Chern-Simons theories. Namely, as we mentioned above, the Calabi-Yau fourfolds we studied in the ${\cal N}=2$ context can be viewed as $G_2$ holonomy manifolds fibered over ${\bf R}$. Turning on RR fluxes in IIA, but {\it not} fibering the Calabi-Yau threefold over ${\bf R}$ corresponds to replacing the ${\cal N}=2$ Chern-Simons couplings in sections 5-8 by ${\cal N}=1$ Chern-Simons couplings. The type IIA/M theory duality then relates this to $N$ M2 branes on a
$G_2$ holonomy manifold obtained by fixing a point in the base ${\bf R}$ of the corresponding Calabi-Yau four-fold $X_4$. Moreover, the RR 2-form fluxes are geometrized in M-theory, but RR 4-form fluxes become $G$-fluxes on the $G_2$ manifold.

String duality now implies that, in the IR, the theory on $N$ D2 branes in these IIA backgrounds becomes the theory on $N$ M2 branes probing the corresponding $G_2$ holonomy manifolds. Since the gauge kinetic terms vanish in this limit, the theory should naturally become the ${\cal N}=1$ quiver Chern-Simons theory. This provides a Lagrangian description of the low energy theory on the $N$ M2 branes probing a $G_2$ holonomy manifold. Note that there are no known Einstein manifolds which are cones over $G_2$ holonomy manifolds times ${\bf R}$ \acharya , so we don't expect these theories to flow to conformal field theories in three dimensions. It will be interesting to see if some insights into the dynamics
of these theories could still be obtained.

\vskip 1cm

\centerline{\bf Acknowledgments}

We would like to thank C. Beem and Y. Nakayama for fruitful discussions, and collaboration on an earlier related project. We are also grateful to O. Ganor for valuable discussions. We thank C. Vafa for colaboration on \AV\ which was crucial for this paper.
The author is indebted to A. Hanany for a beautiful talk at BCTP which initiated this work, and discussions at the early
stage of the project.

The research of M.A. is supported in part by the UC Berkeley Center for
Theoretical Physics and the
NSF grant PHY-0457317.

\listrefs
\end